\documentclass[reprint,amsmath,amssymb,aps]{revtex4-2}

\usepackage{graphicx}
\usepackage{subfigure}

\begin{document}


\title{Free energy analyses of cell-penetrating peptides using the weighted ensemble method}



\author{Seungho Choe}
\email{schoe@dgist.ac.kr}
\affiliation{Department of Energy Science \& Engineering, Daegu Gyeongbuk Institute of Science \& Technolgy (DGIST), Daegu 42988, South Korea}
\affiliation{Energy Science \& Engineering Research Center, Daegu Gyeongbuk Institute of Science \& Technolgy (DGIST), Daegu 42988, South Korea}



\begin{abstract}
Cell-penetrating peptides (CPPs) have been widely used for drug-delivery agents; however, it has not been fully understood how they translocate across cell membranes. The Weighted Ensemble (WE) method, one of powerful and flexible path sampling techniques, can be helpful to reveal translocation paths and free energy barriers along those paths. Within the WE approach we show how Arg$_9$ (nona-arginine) and Tat interact with a DOPC/DOPG (4:1) model membrane, and we present free energy (or potential mean of forces, PMFs) profiles  of penetration, although a translocation across the membrane has not been observed in the current simulations. Two different compositions of lipid molecules were also tried and compared. Our approach can be applied to any CPPs interacting with various model membranes, and it will provide useful information regarding the transport mechanisms of CPPs.
\end{abstract}


\maketitle




\section{Introduction}

Cell-penetrating peptides(CPPs) have been extensively studied for a long time since they are capable of
transporting various cargoes(e.g., proteins, peptides, DNAs, and even small drugs) into cells \cite{Guidotti2017}.
Various factors, including the concentration of CPPs and the properties of the membrane affect the transport mechanisms of CPPs  \cite{Pisa2014,Ruseska2020}.
It has been known that the translocation
mechanisms of different families of CPPs are not the same, and most CPPs can have more than a single
pathway depending on the experimental conditions such as a concentration of CPPs \cite{Madani2011}.

Arginine (R)-rich peptides have been extensively studied because of their effectiveness in translocation \cite{Fretz07,
Ruseska2020}.
One of the possible scenarios for the
 insertion of R-rich peptides into the lipid bilayer is a strong interaction with negatively charged phospholipid heads.

Molecular dynamics (MD) simulations have been used to investigate functional properties of various CPPs and their interactions with many different lipids; however, the mechanism of translocation of CPPs and interactions with lipids
are still under debate. Herce et al. \cite{Herce2007,Herce2009} showed in their MD simulations that
the attractive interactions between the Arg$_9$ (or Tat) peptides and the phosphate groups of the phospholipids results in significant local distortions of the bilayer, and these distortions lead to the formation of a toroidal pore.
However, Herce et al.'s simulation was criticized by Yesylevskyy et al. \cite{Yesylevskyy2009} that the spontaneous translocation of CPPs in MD simulations is not expected within a short time scale (100 $\sim$ 200 ns). It is believed that the time required for the translocation of CPP is on the order of minutes \cite{Zorko2005}.

It is unclear that we can observe the translocation of CPPs across the membranes using conventional all-atom MD simulations. However, an MD simulation is still one of valuable tools to study protein-lipid interactions, and it can provide us with detailed information such as the contribution of electrostatic energy between CPPs and the membranes, effects of water molecules, etc. In our previous work \cite{Choe2020}, we found that the electrostatic interaction between Arg$_9$ and a DOPC/DOPG(4:1) membrane plays a role at the initial stage of translocation. We also quantitatively showed that a number of water molecules coordinated by Arg$_9$ was increased during the penetration, and this led to a membrane thinning and a decreased bending rigidity of the membrane.  However, we couldn't find a translocation of Arg$_9$ across the membrane within a 1 $\mu s$ time scale.

Due to the short time scale achieved in current MD simulations, people have been using biased simulations (e.g., umbrella sampling \cite{Pourmousa2013,Tesei2017,Choong2021}, steered MD simulations \cite{Akhunzada2017}) to study CPPs and their interactions with membranes. The umbrella sampling is very popular to obtain free energy barrier between CPPs and membranes. However, people have noticed that there could be an artifact in the free energy analysis due to a short simulation time in each window.

In this study, we implement a weighted ensemble (WE) method \cite{Zuckerman2017} in our MD simulations (see more in the Methods section). The WE method is one of very flexible path sampling techniques and it is easy to implement. A detailed review was given by Zuckerman and Chong \cite{Zuckerman2017}. We use the WESTPA software \cite{westpa,Bogetti2019}, which has been widely applied to various systems, ranging from atomistic to cellular scale. It turns out that the WE method is very effective for studying interactions between CPPs and membranes and for getting free energy barriers in the current study.

First, we apply the WE method to Arg$_9$ with a DOPC/DOPG(4:1) membrane used in the previous study \cite{Choe2020} as well as Tat with identical lipids. Only a few simulations used mixtures of lipids(e.g., a mixture of
DOPC/DOPG or DOPC/DOPE) so far, and it is still unclear whether the surface charge of a membrane affects  the translocation of CPPs.
For example,
a DOPC/DOPG(1:1) membrane appeared the better host for the translocation of KR9C in experiments \cite{Crosio2019} and
the fluorescence probe carboxyfluorescein-labeled R9 (CF-R9) translocates continuously across the lipid membrane of single
DOPG/DOPC (2/8), DOPG/DOPC (4/6), and DLPG/
DTPC (2/8) giant unilamellar vesicles (GUVs) and enters these GUVs without pore
formation \cite{Sharmin2016}. However, Herce et al. showed in their experiments that DOPG lipids are not necessary for the Arg$_9$ peptides to penetrate bilayers \cite{Herce2009}.
Thus, we also simulate two different lipid compositions,  DOPC/DOPE(4:1) and DOPC lipids, respectively, to see if the surface charge of model membranes affects the penetration of CPPs. Then, the free energy barriers between CPPs and model membranes can be obtained using the WE simulation data, and these free energy calculations will be helpful to understand the transport mechanisms of CPPs.

\section{Methods \label{sec_methods}}

\subsection{Equilibrium simulations}

All simulations were performed using the NAMD package \cite{namd} and CHARMM36 force field \cite{charmm}.
The following systems were well equilibrated before starting the WE simulations:  4 Arg$_9$ with a DOPC/DOPG(4:1) membrane (System I), 4 Tat with a DOPC/DOPG(4:1) membrane (System II), 4 Arg$_9$ with a DOPC/DOPE(4:1) membrane (System III), and 4 Arg$_9$ with a DOPC membrane (System IV).
CHARMM-GUI \cite{charmm-gui} was used to setup a DOPC/DOPG(4:1) membrane, a DOPC/DOPE(4:1) membrane, a DOPC membrane and TIP3P water molecules. The DOPC/DOPG(4:1) mixture consists of 76 DOPC and 19 DOPG lipids in each layer in System I.
The same model membrane was used in System II. System III has a DOPC/DOPE(4:1) mixture which consists of 80 DOPC and 20 DOPE lipids in each layer. System IV has 95 DOPC lipids in each layer.
K and Cl ions were added in each system to make a concentration of 150 mM.
All Arg$_9$ (in System I, III, and IV) or Tat (in System II) peptides were
initially located in the upper solution, and they were bound to the upper layer during the equilibration.

The NPT simulations were
performed at T = 310K. Temperature and pressure were kept constant using Langevin
dynamics. An external electric field(0.05 V/nm) was applied in the negative z-direction(from CPPs
to the membrane) as suggested in previous work \cite{Herce2009,Walrant2012} and also in our previous simulations \cite{Choe2020} to account for the
transmembrane potential \cite{Roux2008}. The particle-mesh Ewald(PME) algorithm was used to compute the
electric forces, and the SHAKE algorithm was used to allow a 2 fs time step during the whole simulations.

In the case of System I, we use a pre-equilibrated structure in the previous study \cite{Choe2020}, which was equilibrated up to 1 $\mu s$. System II was also equilibrated for at least $1 \mu s$.
The other two systems (System III and IV) were equilibrated for 100 ns before starting the WE simulations. This time scale is enough for CPPs to close to model membranes. Fig. S1 (see Supplementary Information)  shows both an initial setup and a snapshot after equilibration for each system, and Fig. S2 presents radial distribution functions of the followings: phosphorus atoms vs. water, phosphorus atoms vs. oxygen in water, ester oxygen vs. oxygen in water.
These plots show that the quality of hydration of membranes is very similar to each other. During the equilibration, CPPs were confined in the upper water box so that there is no interaction between CPPs and the lower leaflet of model membranes. Whenever a CPP is leaving the upper water box, a small force was applied to pull that CPP inside the box. All CPPs in each system were well contacted with the lipid molecules after the equilibration, and then the WE simulations were performed using those equilibrated systems.

\subsection{Weighted Ensemble (WE) simulations}

We use the WESTPA (The Weighted Ensemble Simulation Toolkit with Parallelization and
Analysis) software package \cite{westpa,Bogetti2019} to enable the simulation of rare events, for example, translocation of CPPs across a model membrane. WESTPA is an open source, and its utility has been proved for a broad range of problems.
All WE trajectories are unbiased and hence used to calculate conditional probabilities or transition rates.

To use the WE method in MD simulations, we need to define a progress coordinate, a number of bins, a number of walkers (child simulations) in each bin, and a time interval for splitting and combining trajectories \cite{westpa}.
We define the progress coordinate as a distance in z-direction between the center of mass of phosphorus atoms in the upper leaflet and that of a CPP (Arg$_9$ or Tat). After equilibration of each system, an initial distance between phosphorus atoms and a CPP was measured, and boundaries were set using this initial position and the position of the center of the membrane, for example, [- 18 \AA ~(the center of membrane), 3 \AA ~(the initial distance)]. Each bin size was 0.25 \AA ~and, the number of walkers in each bin was 5. The time interval for splitting and combining the trajectories was set 5 ps during all the WE simulations. The progress coordinate was calculated using MDAnalysis \cite{Michaud2011}. The potential mean of force (PMF) profiles were obtained from each WE simulation data using "w\_pdist" and "plothist" codes in the WESTPA package \cite{westpa,Bogetti2019}.

\section{Results}

\subsection{WE simulations show much-enhanced penetration within a short amount of time}

We observe that both Arg$_9$ and Tat penetrate through the middle of the model membrane during the WE simulations; however,  translocation across the membrane has not been observed in the current simulations.

\begin{figure*}[ht]	
\includegraphics[width=12cm,height=8cm]{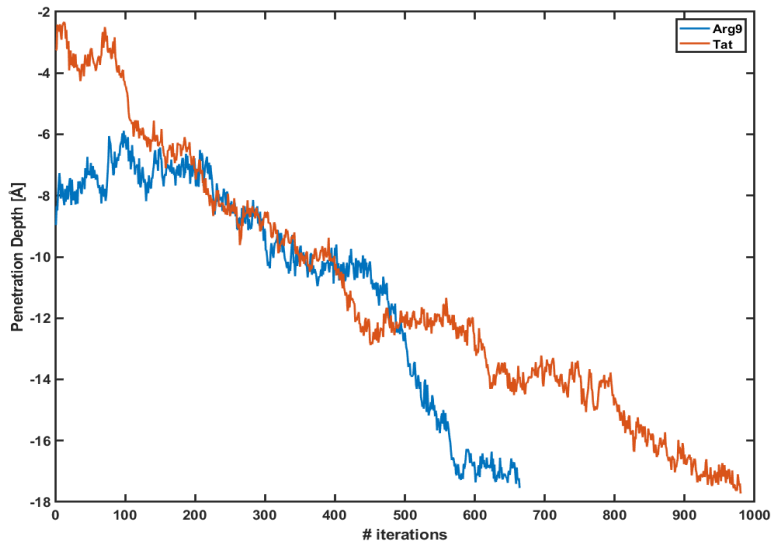}
\caption{The penetration depth of Arg$_9$ (blue line) and Tat(red line) vs. a number of iterations \label{fig1}}
\end{figure*}

Fig. \ref{fig1}
shows the penetration depth of both Arg$_9$ and Tat vs. the number of iterations of each WE simulation.
Here, we define the penetration depth as a distance between the center of mass of CPP and that of phosphorus atoms of the upper leaflet. The penetration depth of Arg$_9$ in the previous equilibrium simulation was about $-6 \sim -7$ \AA \cite{Choe2020}, and the current WE simulation shows $- 17.6$ \AA ~and $-17.7$ \AA ~for Arg$_9$ and Tat, respectively. The initial positions of Arg$_9$ and Tat after equilibration are different from each other; however, they quickly penetrate the DOPC/DOPG(4:1) membrane.
Note that these plots show only one of the trajectories of each simulation which shows the maximum penetration depth. Multiple trajectories can not reach the middle of the membrane.
 As one can see in the figure, the penetration depth changes a lot within a very short time scale. As described in the Methods section, each iteration corresponds to 5 ps and the figure shows that both CPPs reach their maximum penetration depth within a very short time scale (3 $\sim$ 5 ns) after the WE simulation started.
Therefore, the WE method provides us with a very effective tool to overcome potential energy barriers.

\begin{figure*}[ht]	
\includegraphics[width=6cm,height=4cm]{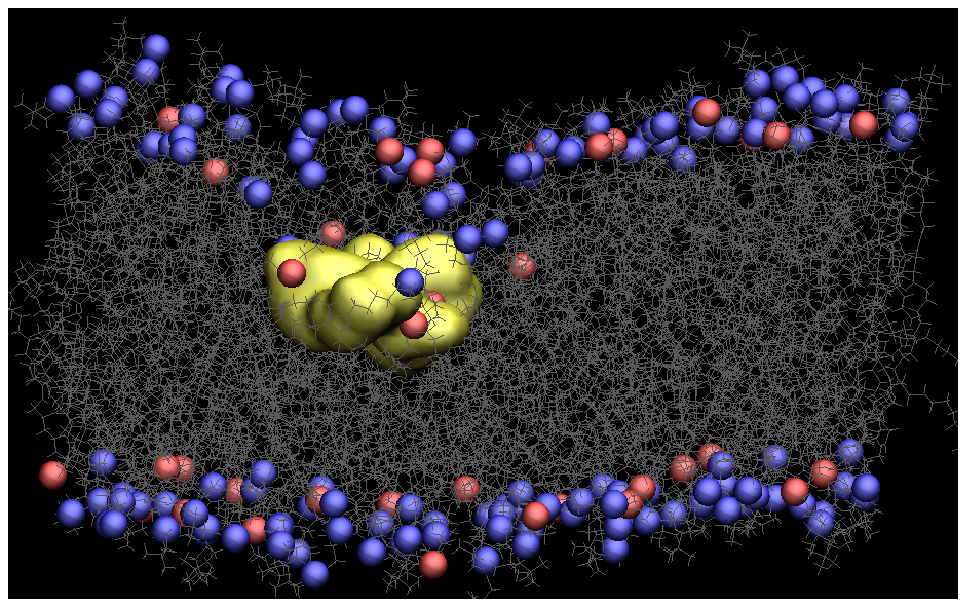}
\includegraphics[width=6cm,height=4cm]{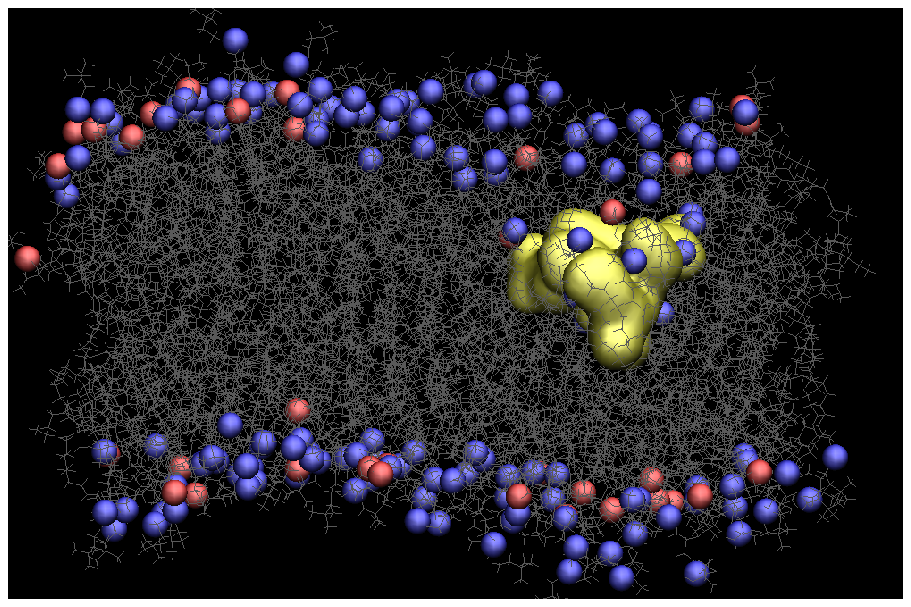}
\caption{Snapshots of Arg$_9$ (left) and Tat(right) with a DOPC/DOPG(4:1) model membrane (yellow: CPP, gray: lipids, blue and red: phosphorus atoms of DOPC and DOPG, respectively). \label{fig2}}
\end{figure*}

Fig. \ref{fig2} presents snapshots of both Arg$_9$ and Tat, which show the maximum penetration depth during the WE simulation. The yellow shows each CPP, and gray is the lipids. The blue and red are phosphorus atoms of the upper and the lower leaflets. Water molecules, ions,  and the other three CPPs are omitted for clarity. As one can see in Fig. \ref{fig2}, the
membrane is deformed a lot when Arg$_9$ or Tat penetrates the middle of the membrane. Fig. S3 in Supplementary Information shows membrane curvature more clearly. As shown in the previous simulation \cite{Choe2020}, a number of water molecules around a CPP is increased when the CPP penetrates the membrane, and a membrane thinning is also observed in the current WE simulations.

\subsection{Electrostatic energy between a CPP and a model membrane plays a role during the translocation}

In our previous equilibrium  simulation of Arg$_9$ with a DOPC/DOPG(4:1) membrane, we showed that electrostatic energy between Arg$_9$ and the membrane is strongly correlated with the penetration depth of Arg$_9$\cite{Choe2020}. This proved that electrostatic interaction between a CPP and a membrane is essential at the early translocation stage \cite{Galassi2021}.

We calculate the same electrostatic energy between Arg$_9$ (or Tat) and the membrane. We want to see if there is still a strong correlation between  electrostatic energy and  penetration depth during the WE simulations.
We use two trajectories in Fig. \ref{fig1} to calculate electrostatic energy.

\begin{figure*}[ht]	
\includegraphics[width=12cm,height=8cm]{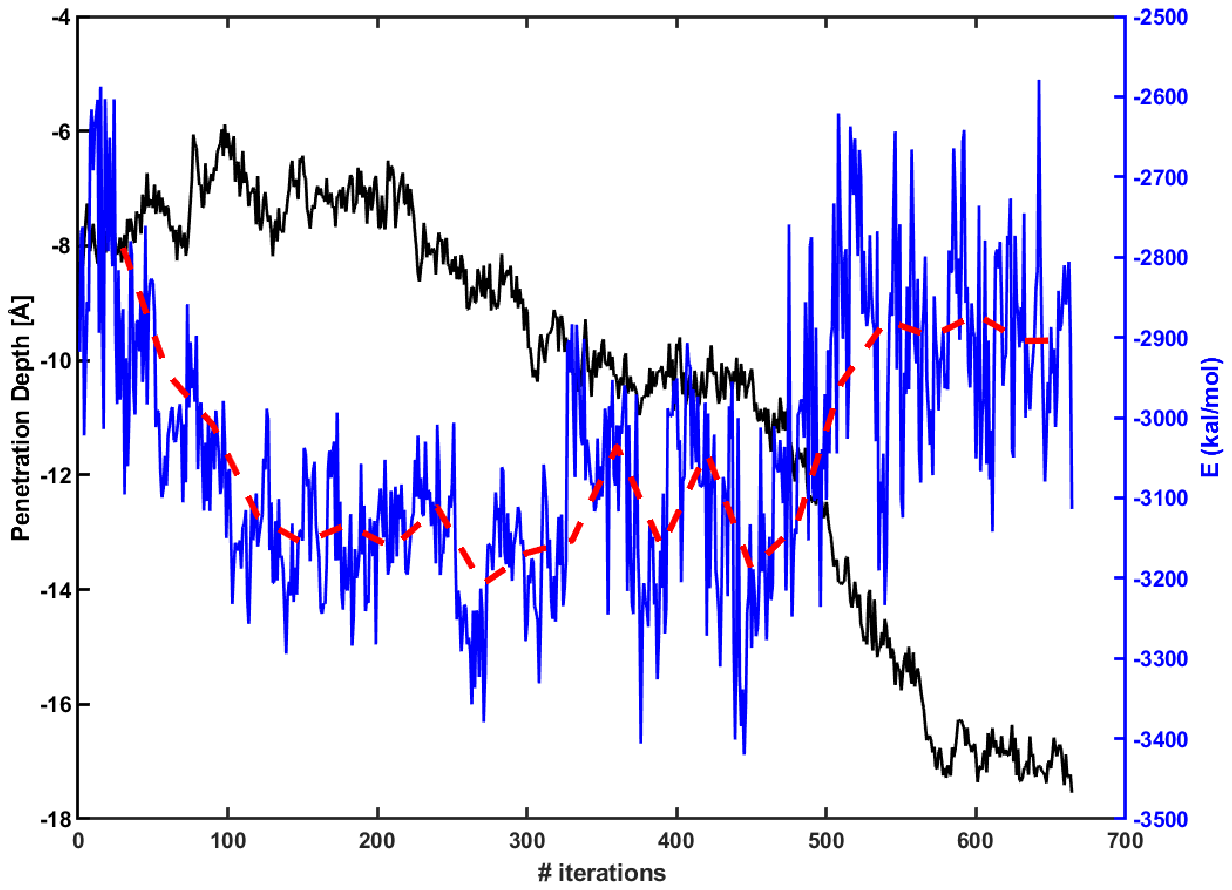}
\caption{The penetration depth of Arg$_9$ (left axis, black line) vs. electrostatic energy between Arg$_9$ and lipids molecules within 15 \AA ~of Arg$_9$ (right axis, blue line). The red dotted line is an average over every 30 iterations.  \label{fig3}}
\end{figure*}

Fig. \ref{fig3} presents the penetration depth (left axis, black line) vs. electrostatic energy (right axis, blue line) between Arg$_9$ and the lipid molecules within 15 \AA ~of Arg$_9$. The red dotted line is an average over every 30 iterations.
The figure shows that an absolute magnitude of electrostatic energy becomes smaller when Arg$_9$ rapidly penetrates inside the membrane (e.g., between 450th $\sim$ 550th iterations). Therefore, one can expect that the penetration increases as the magnitude of electrostatic energy becomes smaller.

\begin{figure*}[ht]	
\includegraphics[width=12cm,height=8cm]{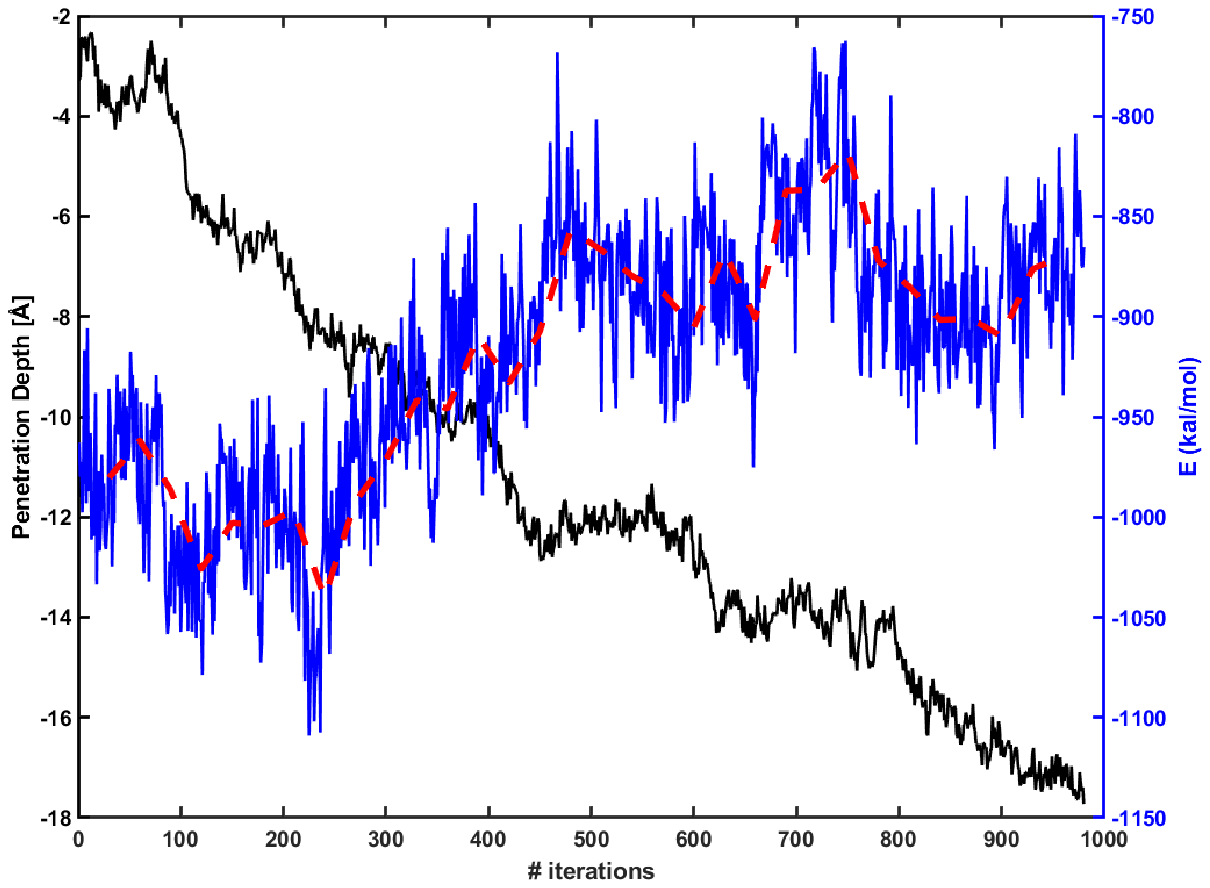}
\caption{Same as in Fig. \ref{fig3} , but for Tat\label{fig4}.}
\end{figure*}

We find similar behavior in the case of Tat.
Fig. \ref{fig4} shows the same energy calculation for Tat. The red dotted line is an average over every 30 iterations as before.
The penetration increases gradually as the magnitude of electrostatic energy becomes getting smaller.

At the initial stage of translocation, it is well known that the interaction between positive charges of CPPs and a negative part of lipid molecules is essential for binding.
The current simulations show that electrostatic interaction between CPPs and membranes still plays a role during the translocation of CPPs. In System I \& System II Arg$_9$ and Tat strongly interact with DOPG lipids, and the translocation seems not easy.
However, our simulations show that the translocation of CPPs is possible even in this strong electrostatic interaction. During the translocation, the electrostatic interaction should be diminished.
We conjecture that water molecules play a role to make electrostatic energy weaker. As shown in the previous work \cite{Choe2020}, a number of water molecules around CPPs is increased during the translocation.

\subsection{The free energy profiles based on the WE simulation data show that Arg$_9$ penetrates through a DOPC/DOPG(4:1) model membrane much easier than Tat does }

This section presents the potential mean of force (PMF) profiles of Arg$_9$ and Tat along the penetration paths shown in Fig. \ref{fig1}. Another advantage of using the WE method is that it provides a free energy analysis without any biased sampling (e.g., umbrella sampling) \cite{westpa}. Therefore, these PMF profiles will tell us how much potential energy barriers exist along those paths and which CPP can penetrate the membrane more efficiently.

\begin{figure*}[ht]	
\includegraphics[width=12cm,height=8cm]{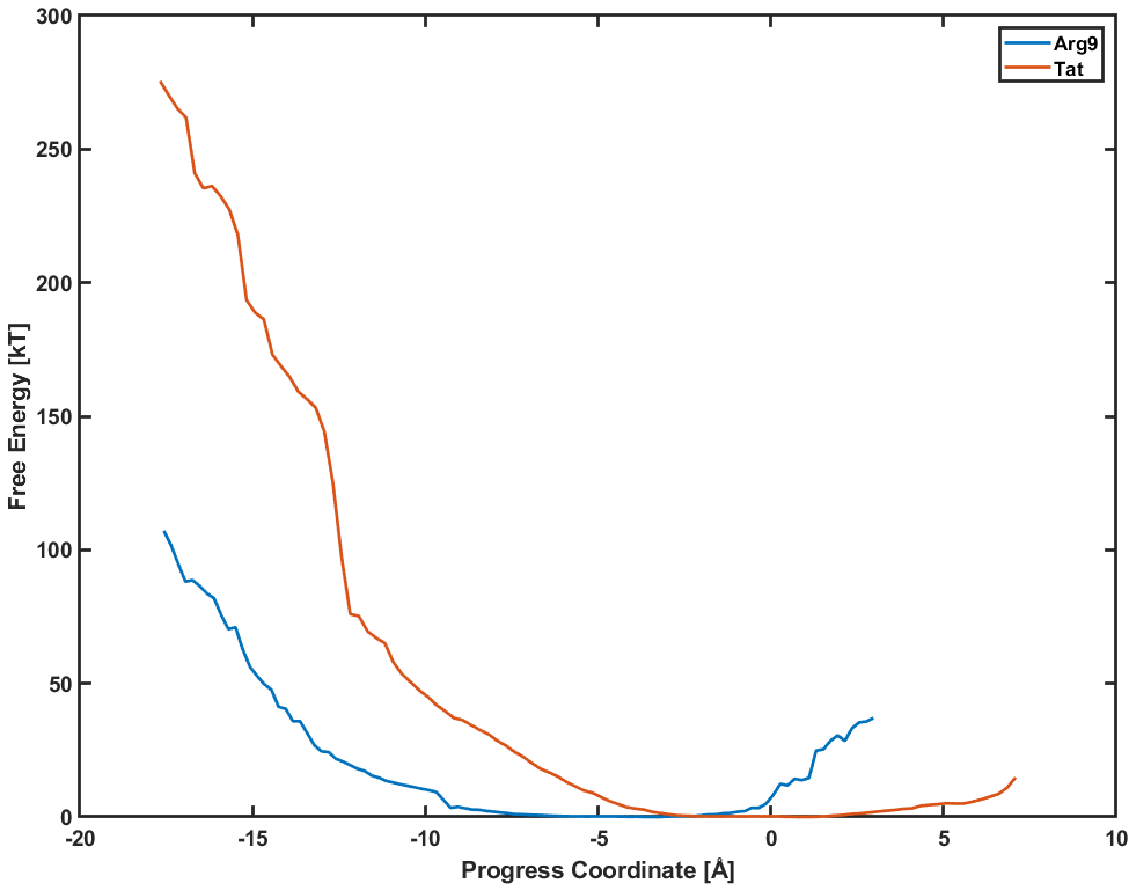}
\caption{Comparison of potential mean of force (PMF) between Arg$_9$(blue line) and Tat(red line).   \label{fig5}}
\end{figure*}

Fig. \ref{fig5} shows the PMF profiles of both Arg$_9$ and Tat as a function of a progress coordinate (or reaction coordinate). We use a penetration depth as the progress coordinate. The progress coordinate becomes negative when Arg$_9$ and Tat penetrate below the upper leaflet of the membrane. The blue line is the PMF profile for Arg$_9$ and the red line for Tat. One can notice that the free energy of Arg$_9$ is much lower than that of Tat.
Fig. \ref{fig5} shows that a free energy barrier is about 100 kT for Arg$_9$ to reach the middle of the membrane while it is about 280 kT for Tat to reach a similar position. This means that Arg$_9$ penetrates the DOPC/DOPG(4:1) membrane much easier than Tat does. However, these analyses can depend on the type of model membranes, and we can not conclude that Arg$_9$ is always more efficient for penetrating membranes than Tat regardless of membrane types.
The order of magnitude of free energy for Arg$_9$ ($\sim$ 100 kT) is very similar to a previous result \cite{Huang2013}, which used umbrella samplings. On the other hand, Tat's  order of magnitude of free energy is much higher than those of previous simulations \cite{Yesylevskyy2009,Choong2021}. Note that most of the previous free energy analyses were done with a single lipid molecule (e.g., DOPC or DPPC), while we use a mixed composition of lipid molecules (DOPC/DOPG(4:1)) in our simulations.
Although the magnitude of free energy can depend on systems studied, the observed tendency between Arg$_9$ and Tat is consistent with experimental results \cite{Qian2016,Guterstam2009} which showed more efficient cellular uptake of Arg$_9$ than Tat.

\subsection{Different lipid compositions slightly affect the PMF profiles of Arg$_9$}

\begin{figure*}[ht]	
\includegraphics[width=12cm,height=8cm]{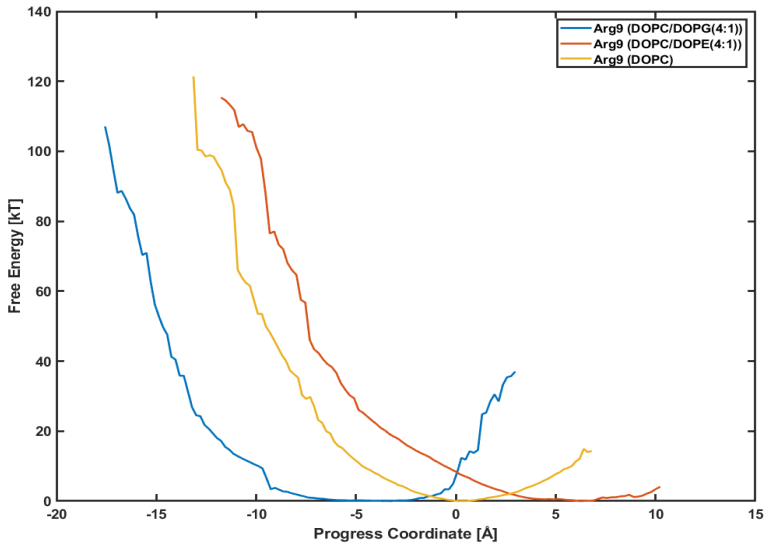}
\caption{Comparison of potential mean of force (PMF). Arg$_9$ with a DOPC/DOPG(4:1) (blue line),  a DOPC/DOPE(4:1) (red line), and a DOPC (orange line), respectively.  \label{fig6}}
\end{figure*}

It is interesting to see whether a different lipid composition can affect the penetration depth and the PMF profile.
Fig. \ref{fig6} presents the PMF profiles of Arg$_9$ with DOPC/DOPG(4:1) lipids, DOPC/DOPE(4:1) lipids, and  DOPC lipids, respectively. The blue line is for Arg$_9$ with DOPC/DOPG(4:1) lipids shown already in Fig. \ref{fig5}. The red one is for Arg$_9$ with DOPC/DOPE(4:1) lipids and the orange for Arg$_9$ with DOPC lipids.
DOPC and DOPE lipids are neutral, while DOPG lipids are negatively charged. We want to see if the translocation of CPPs strongly correlates with the lipid types, e.g., the charge of membranes.

The minimum point in each plot is different from each other because the initial position of each CPP after equilibration is not the same.
Although the initial positions are different from each other, the slope of each plot looks very similar to each other. Although Arg$_9$ in System III (with DOPC/DOPE(4:1) lipids) and Arg$_9$ in System IV (with DOPC lipids) have not reached the final position as of System I (with DOPC/DOPG(4:1)), the similar slope  means similar variations in free energy. Based on these plots, we can conjecture that the penetration of Arg$_9$ is not much dependent on the lipid composition, and the effect of surface charge is minimal.

\section{Discussion}

Both Arg$_9$ and Tat don't show any translocation across the model membranes during the WE simulations. However, the WE method can be helpful to identify the transport mechanisms of CPPs and interactions between CPPs and lipid molecules because conventional equilibrium MD simulations have difficulty in sampling.
Free energy (potential mean of force, PMF) profiles can be obtained without any biased simulation (e.g., umbrella sampling) within the WE approach. The WE simulation and its free energy analyses can be used to study the efficiency of penetration of CPPs.

Is it possible to observe the translocation if we simulate much longer? Currently, both Arg9 and Tat are stuck in the hydrophobic core. They can stay in the hydrophobic core for a long time even if we simulate much longer. They can move up and down a little, but it is not likely to move in either direction, moving back or forward. A visual inspection shows that they are trapped inside the hydrophobic core. The lipids make a small cage around CPPs so that CPPs cannot move easily. It seems that a single CPP cannot cause further deformation of the membrane at the current stage, and thus it is difficult to move.
There is a possibility that the free energy barriers along the downward direction are much higher than those along the upward direction. In that case, it is easier for CPPs to go back to the hydrophilic region where they started. It will be interesting to try the WE simulations from the current positions of both Arg9 and Tat to the hydrophilic area outside, i.e., the positive z-direction. Then, we will obtain another free energy profiles for both Arg9 and Tat, and these profiles will give us a clue.

There are still some limitations in our WE simulations.
First, a time scale of each trajectory shown in Fig.  \ref{fig1} is less than 10 ns.
Note that the total simulation time for each CPP simulation is about 5 $\sim$ 10 $\mu s$, which includes all the walkers (child simulations) in each bin and all the iterations.
This time scale of a single trajectory might be too short to observe a water pore mentioned in the previous results \cite{Yesylevskyy2009,Huang2013,Islam2018,Choong2021}. Second, our PMF profiles are incomplete because we don't find any translocation of CPPs and PMF profiles spanned only from the top to the middle of the model membranes.
It has been known that free energy barriers between CPPs and membranes are changed in the presence of a water pore. For example, Huang et al.\cite{Huang2013} reported that the free energy becomes lower in the presence of a water pore.
This confirms the previous results from MD simulations and a continuum modeling which calculate the free energy of a single arginine residue for translocating across a membrane
\cite{Freites2005,Dorairaj2007,MacCallum2007,Choe2008}.
We need more WE simulation data points to complete the free energy profiles, which include translocation of CPP across the membranes and the effect of the water pore. Third, we haven't attached any cargo to Arg$_9$ (or Tat) in the current study. We expect that the conformational changes both in CPP and the model membranes depend on a type of cargo. CPPs are commonly used with attaching negatively charged molecules and thus the total charge of the system decreases or gets closer to neutral. It is not clear whether the overall charge of the whole system or only the charge of CPP is essential for the internalization. Most electronic neutral CPPs are usually far less powerful than cationic or amphiphilic CPPs. However, the electronic neutral cyclic peptide cyclosporin A (CsA) can have stronger (5-fold or 19-fold) penetrating capacity than other neutral peptides and its penetrating capacity was as strong as Tat \cite{Gao2017}. On the other hand, in the case of oligoarginine ($R_n$: n=4,8,12,16) $R_{12}$ and $R_{16}$ showed more efficient uptake than $R_4$ and $R_8$ \cite{Kosuge2008}.
If only the charge of CPP is essential during the translocation, we can compare the free energy barriers of these oligoarginine within the WE method and identify their penetration efficiency.

It has been known that it is difficult for a single CPP to make a pore and to make a translocation across the membrane \cite{Yesylevskyy2009}.
One of the key factors which makes the water pore is a concentration of CPPs. In our WE simulations, the P(protein)-L(lipids) ratio was 0.042 or 0.05, and these concentrations maybe not be enough to observe the translocation across the membrane.
Another critical factor for translocating CPPs is pH dependence. Experimental results showed the uptake of CPPs depends on pH values\cite{Sun2014,Ouahab2014}, and the uptake of CPPs is enhanced at low pH. It is worth trying MD simulations at low pH to see if any changes  in conformational changes both in CPPs and in the model membranes.
There has been significant progress in the development of constant pH molecular dynamics (pHMD) techniques \cite{Chen2014,Radak2017} and this approach with the WE method would be good to reveal the pH dependence in interactions between CPPs and the model membranes.

Our simulations show characteristics of both the carpet-like model \cite{Pouny92} and the membrane thinning model \cite{Lee05}.
It is interesting to note that an orientation (a direction from the N-terminal to the C-terminal) of CPPs (Arg$_9$ or Tat) in our simulations is  almost parallel to the membrane, and its orientation is changed a little during the entire simulation. This could be one of the reasons that it is difficult for a single CPP to make a water pore.
In this work, the progress coordinate was set for a single CPP, and it was difficult to observe aggregation of CPPs in a local area. To make a water pore, a collective behavior between CPPs seems to be essential, and a high concentration of CPPs would be needed.
It is worth using an increased concentration of CPPs and implementing a progress coordinate as a distance between the center of mass of  the phosphorus atoms of the upper leaflet and that of more than a single CPP (e.g., two Arg$_9$ or two Tat). However, the penetration of two CPPs seems much slower than using a single CPP \cite{Choe2021}.

%




\begin{acknowledgments}
This work was supported by the Faculty start-up fund from DGIST and by the DGIST R$\&$D Program of the Ministry of Science and ICT (21-BRP-12)

The author would like to thank the DGIST Supercomputing Big Data Center for computing resources and
the National Supercomputing Center with supercomputing resources including technical support (KSC-2021-CRE-0296).
\end{acknowledgments}


%

\bibliography{cpp}


\end{document}



\vspace{1cm}
\title{{\LARGE Supplementary Information} \\
 Free energy analyses of cell-penetrating peptides using the weighted ensemble method}

\author{Seungho Choe}
\email{schoe@dgist.ac.kr}
\affiliation{Department of Energy Science \& Engineering, Daegu Gyeongbuk Institute of Science \& Technolgy (DGIST), Daegu 42988, South Korea}
\affiliation{Energy Science \& Engineering Research Center, Daegu Gyeongbuk Institute of Science \& Technolgy (DGIST), Daegu 42988, South Korea}

%

\maketitle

\newpage
\section{System Setup}

System I and System II were equilibrated for at least 1 $\mu s$, while System III and IV were equilibrated for 100 ns. As for System I, we use a pre-equilibrated structure in the previous study \cite{Choe2020}. Fig. S1 shows snapshots of both the initial  and the final structure after finishing equilibration.
Each row corresponds to each system (e.g., System I, II, III, and IV). The first and second columns present the initial setup of each system (both a side view and a top view, respectively). Four Arg$_9$ (or Tat) were initially placed in the upper water box and stayed in the upper water box during equilibration. The third and fourth columns show the final structure after equilibration (both a side view and a top view). A 100ns-long equilibration may be short to fully equilibrate the system (System III and IV); however, a visual inspection shows that four Arg$_9$ were contacted with the lipid molecules. We also plot radial distribution functions to check the hydration profiles of each system Fig. S2 presents radial distribution functions g(r) (a)  phosphorus vs. water (b) phosphorus vs. oxygen in water (c)  easter oxygen vs. oxygen in water. The radial distribution function of each system is quantitatively very similar to each other, and the structures in System III and IV can be used as initial structures for the WE simulations.

\begin{figure*}[ht]
\label{figS1}
\includegraphics[width=3cm,height=3.5cm]{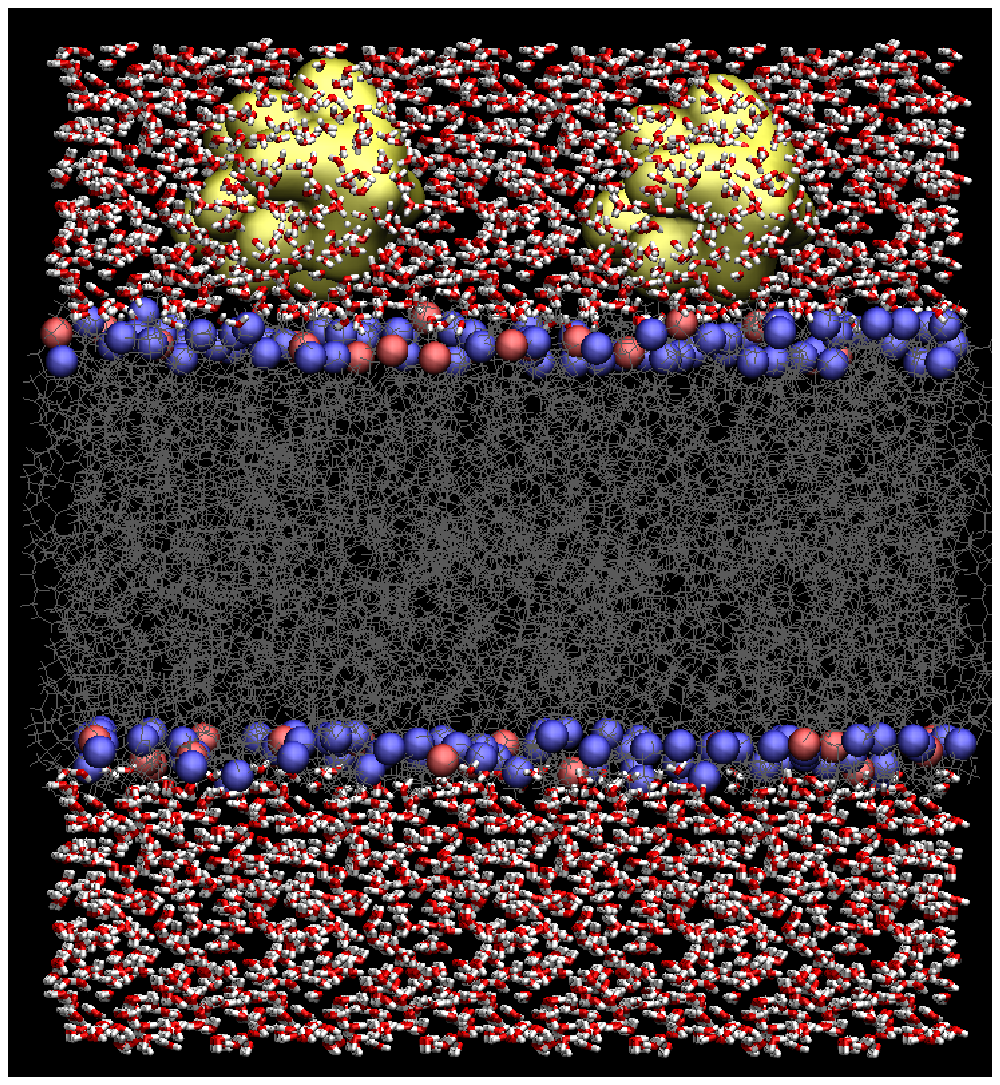}
\includegraphics[width=3cm,height=3.5cm]{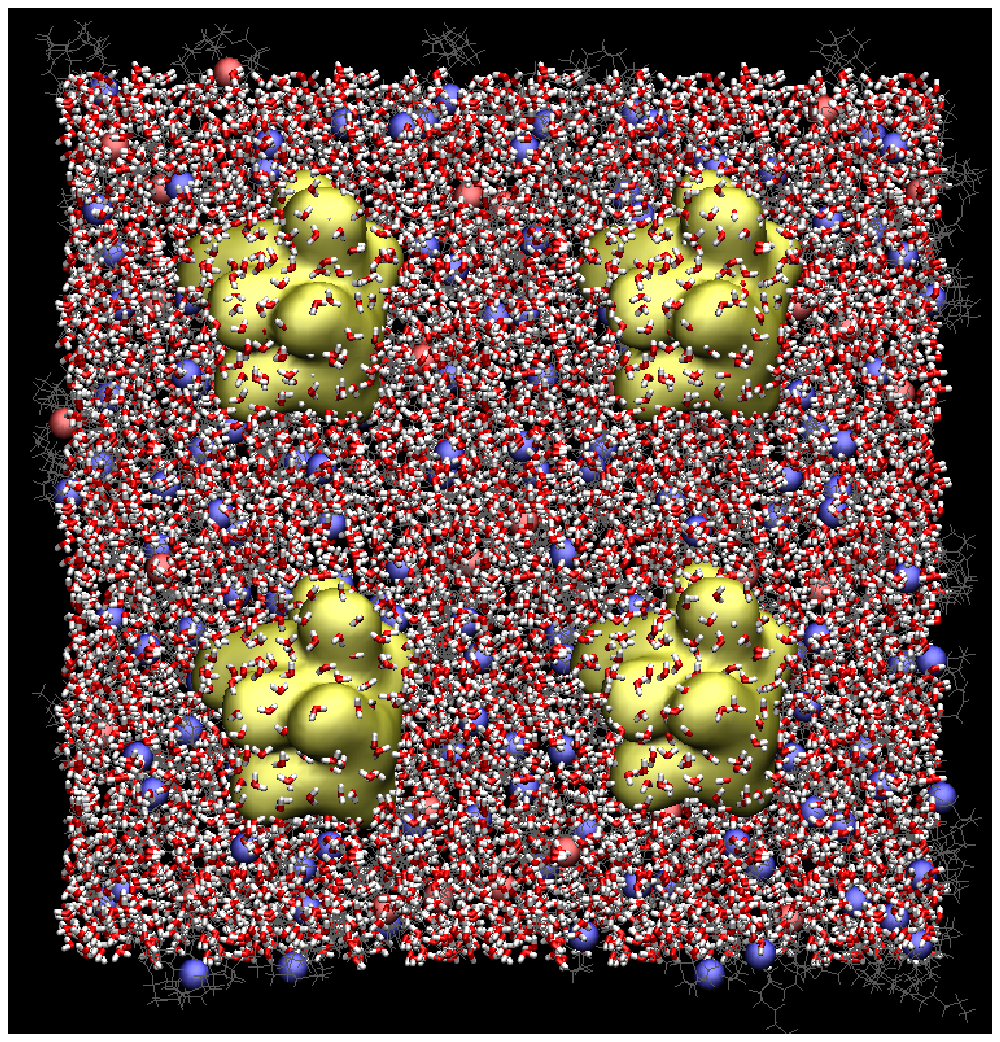}
\includegraphics[width=3cm,height=3.5cm]{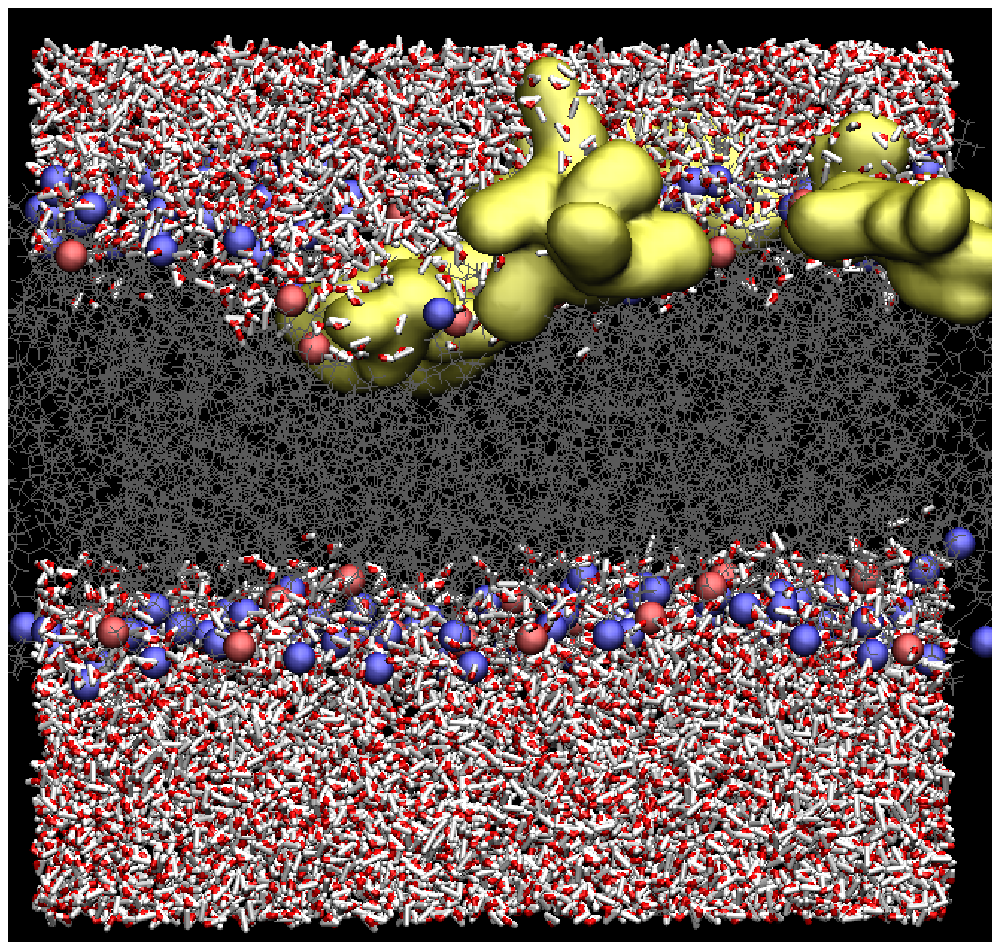}
\includegraphics[width=3cm,height=3.5cm]{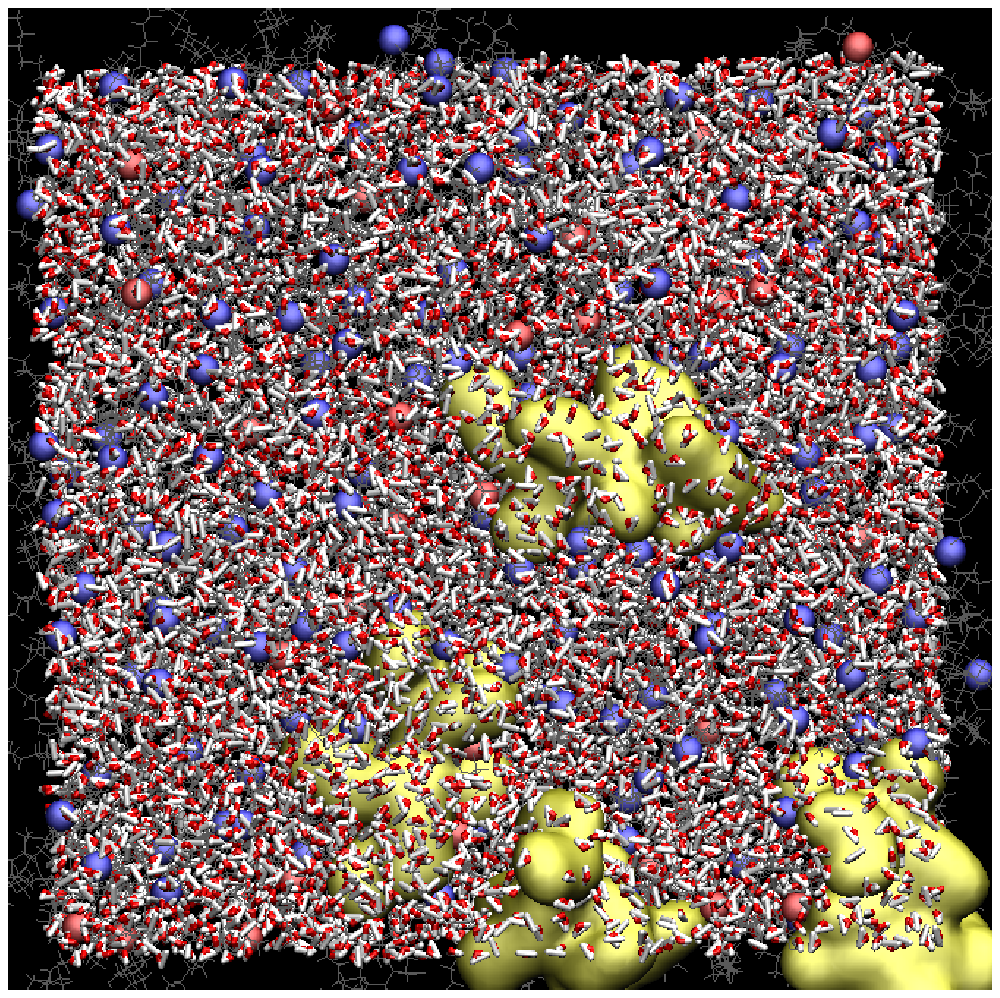}

\includegraphics[width=3cm,height=3.5cm]{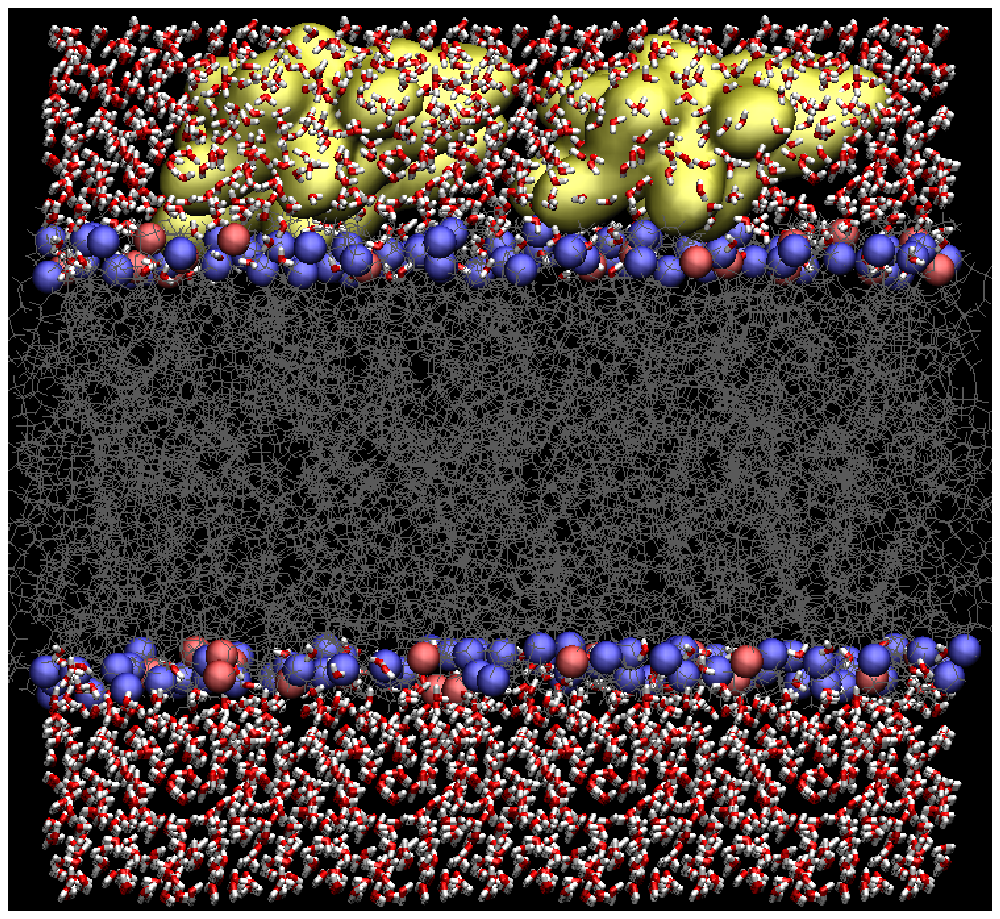}
\includegraphics[width=3cm,height=3.5cm]{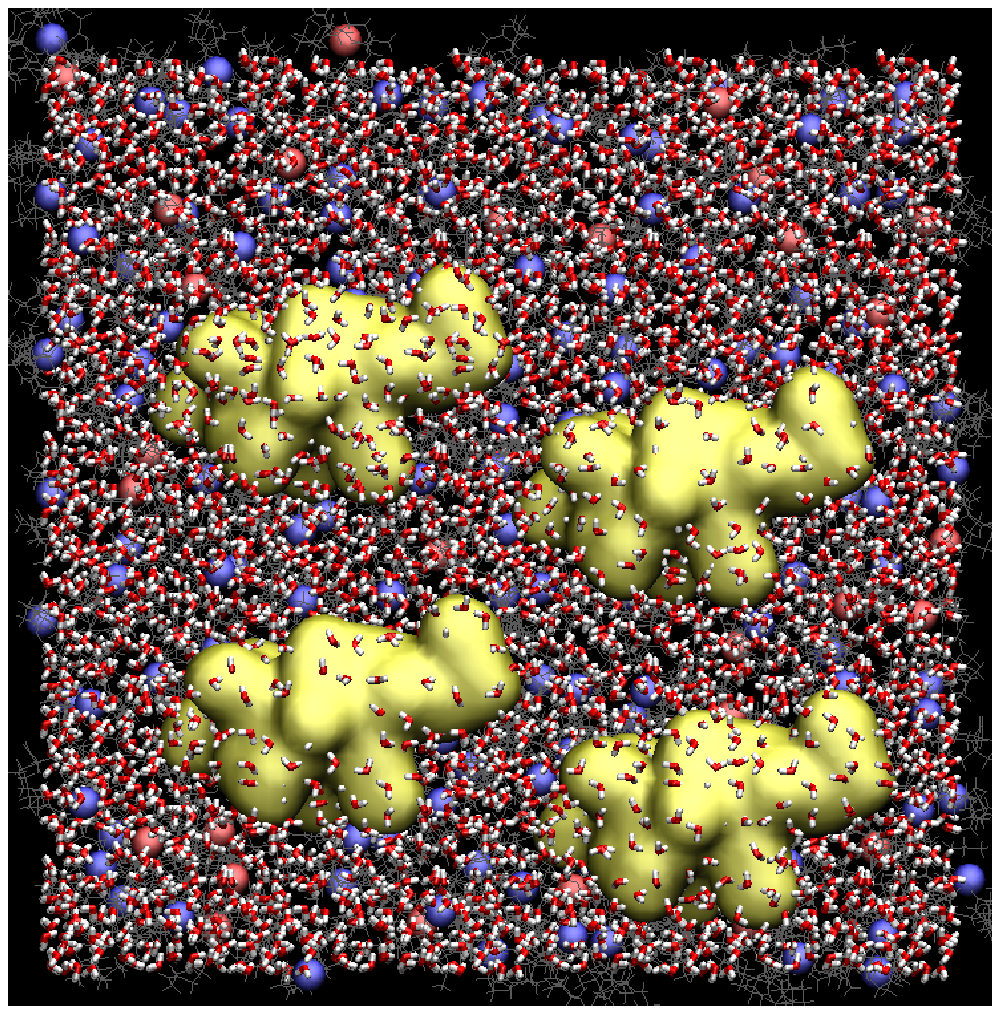}
\includegraphics[width=3cm,height=3.5cm]{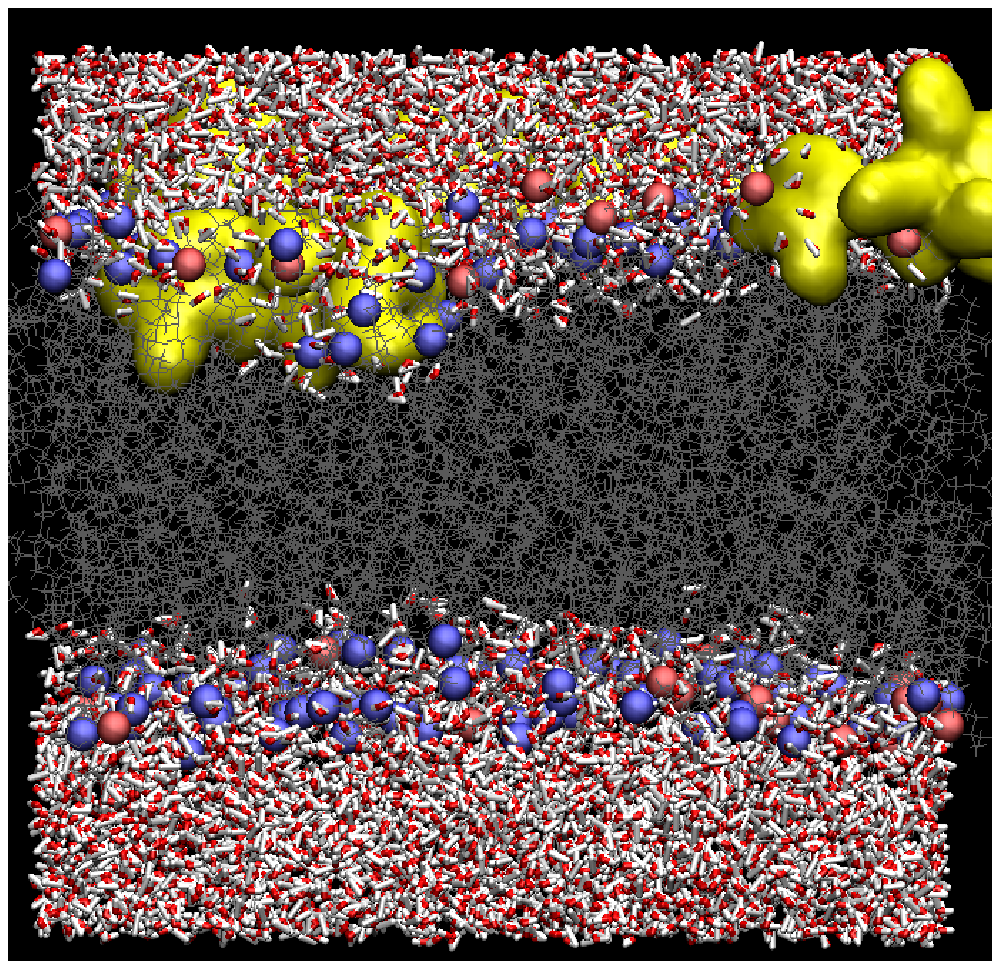}
\includegraphics[width=3cm,height=3.5cm]{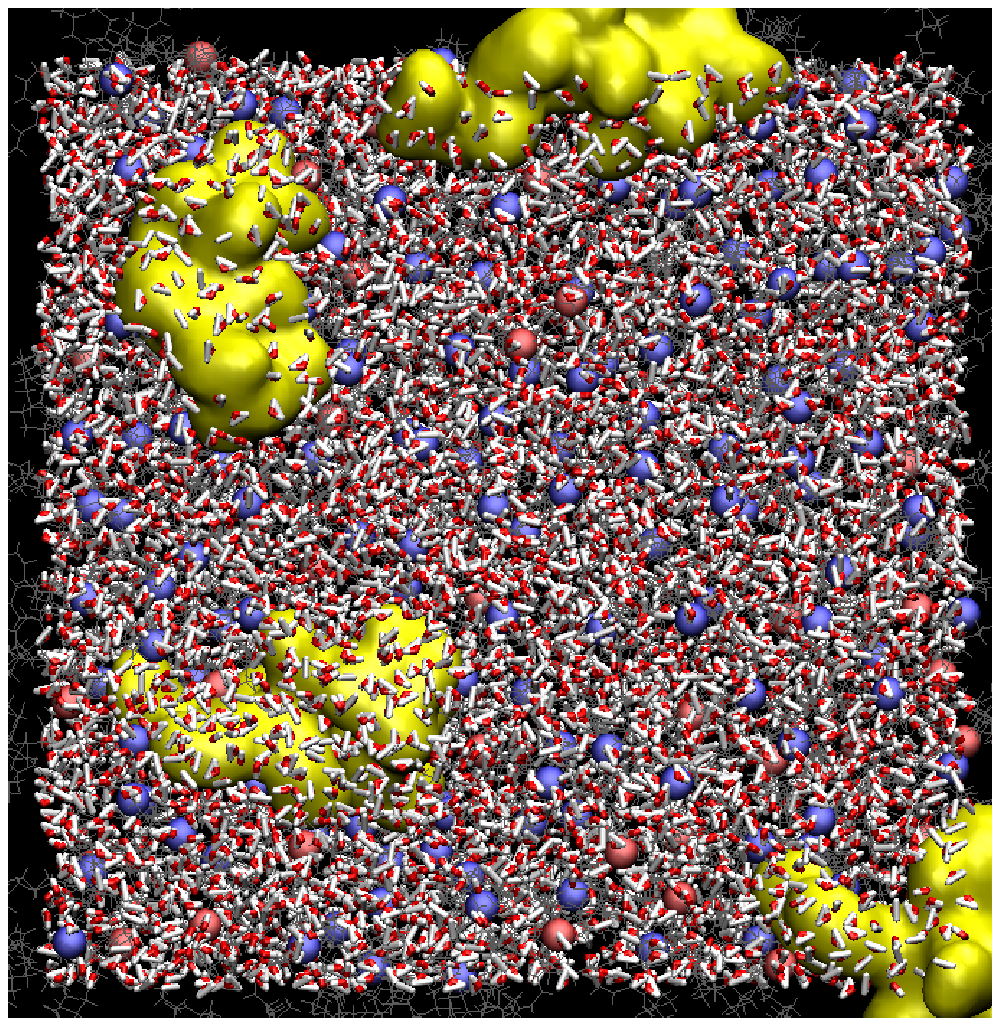}

\includegraphics[width=3cm,height=3.5cm]{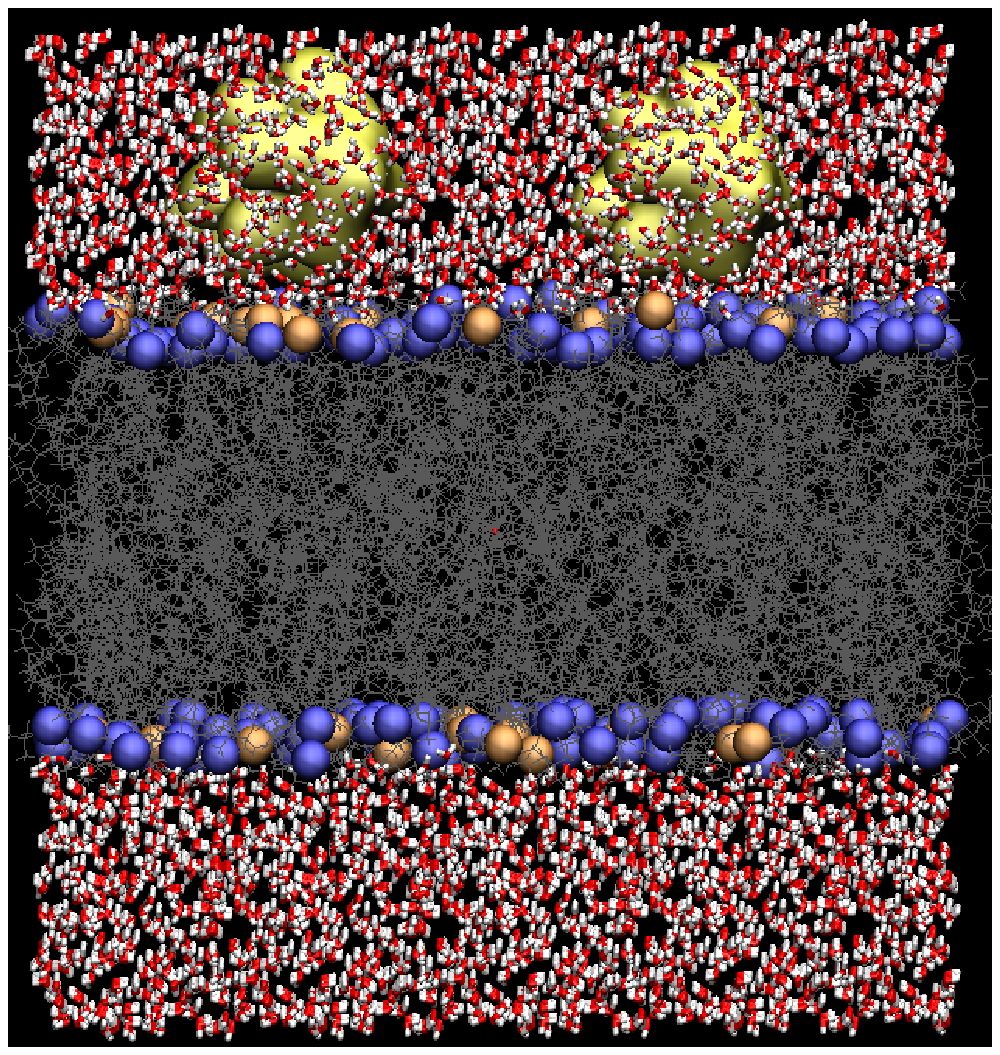}
\includegraphics[width=3cm,height=3.5cm]{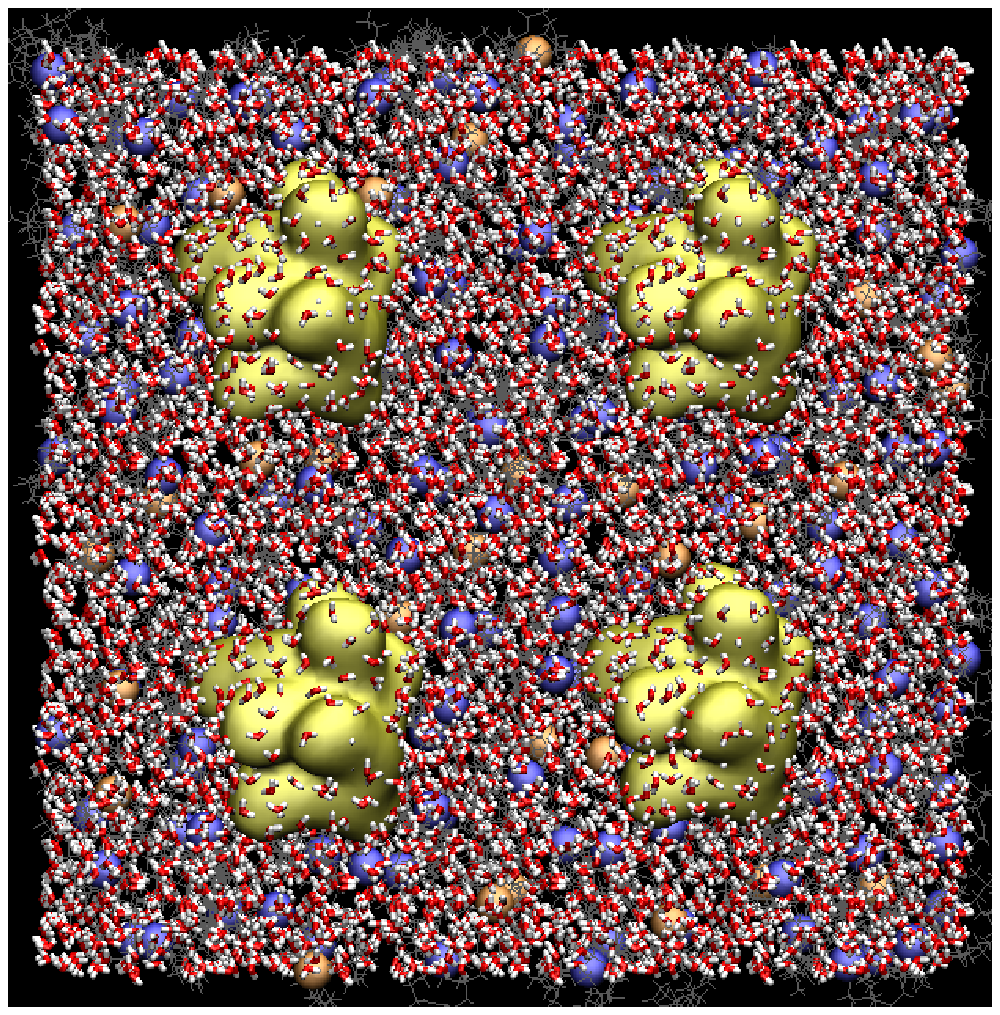}
\includegraphics[width=3cm,height=3.5cm]{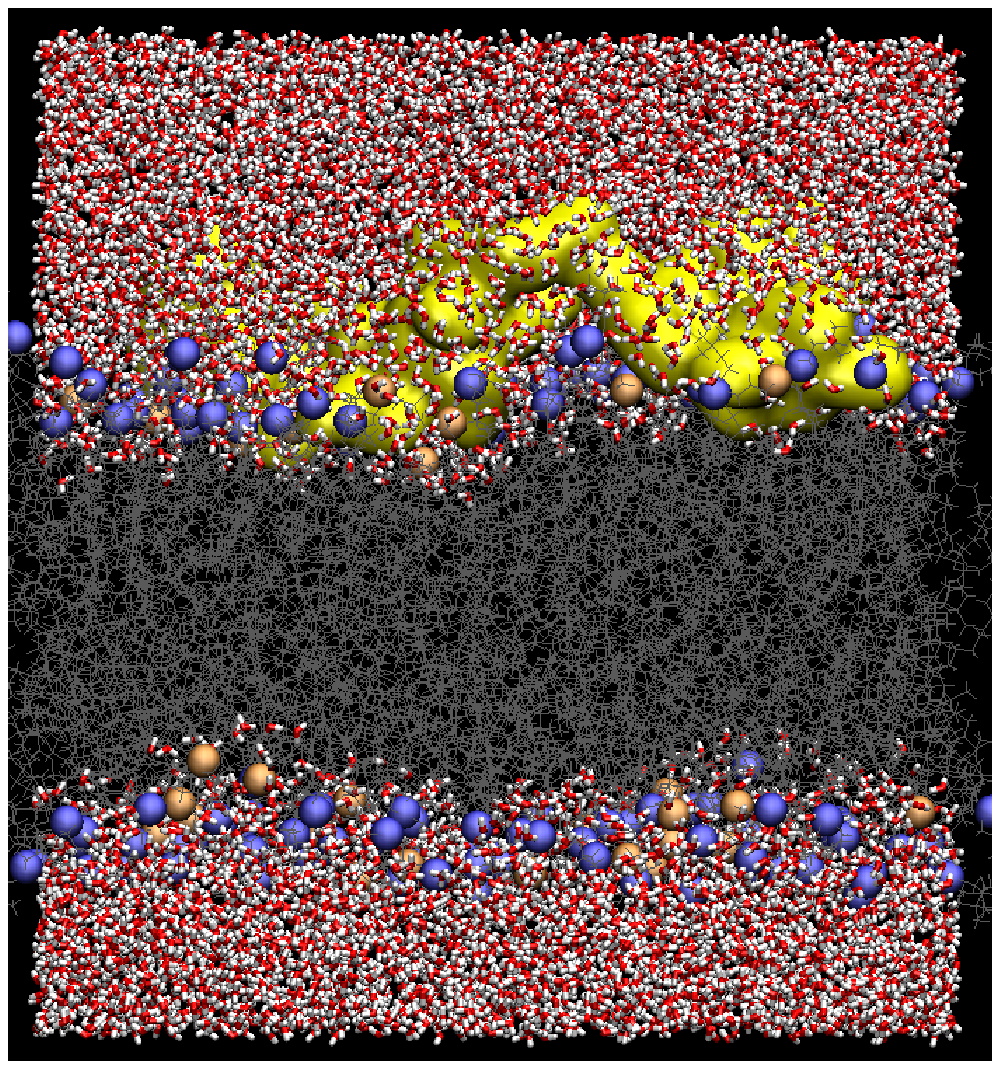}
\includegraphics[width=3cm,height=3.5cm]{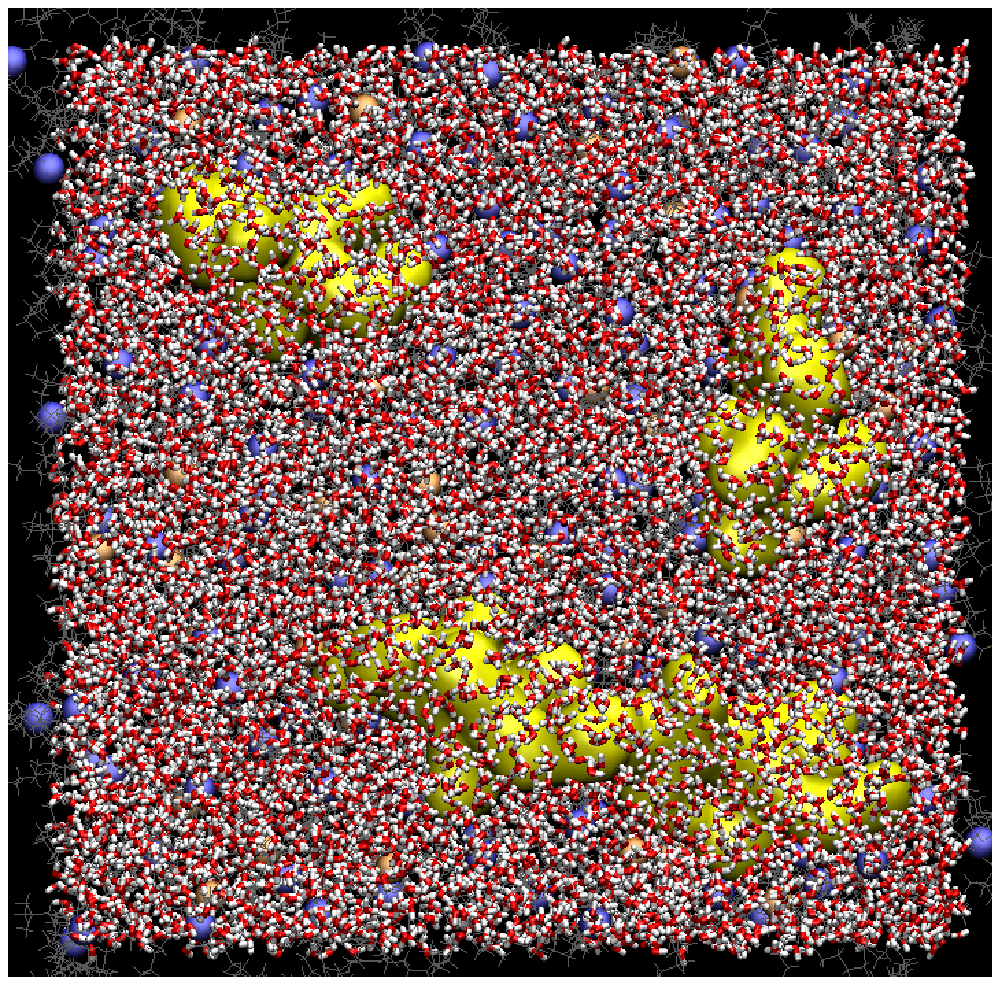}

\includegraphics[width=3cm,height=3.5cm]{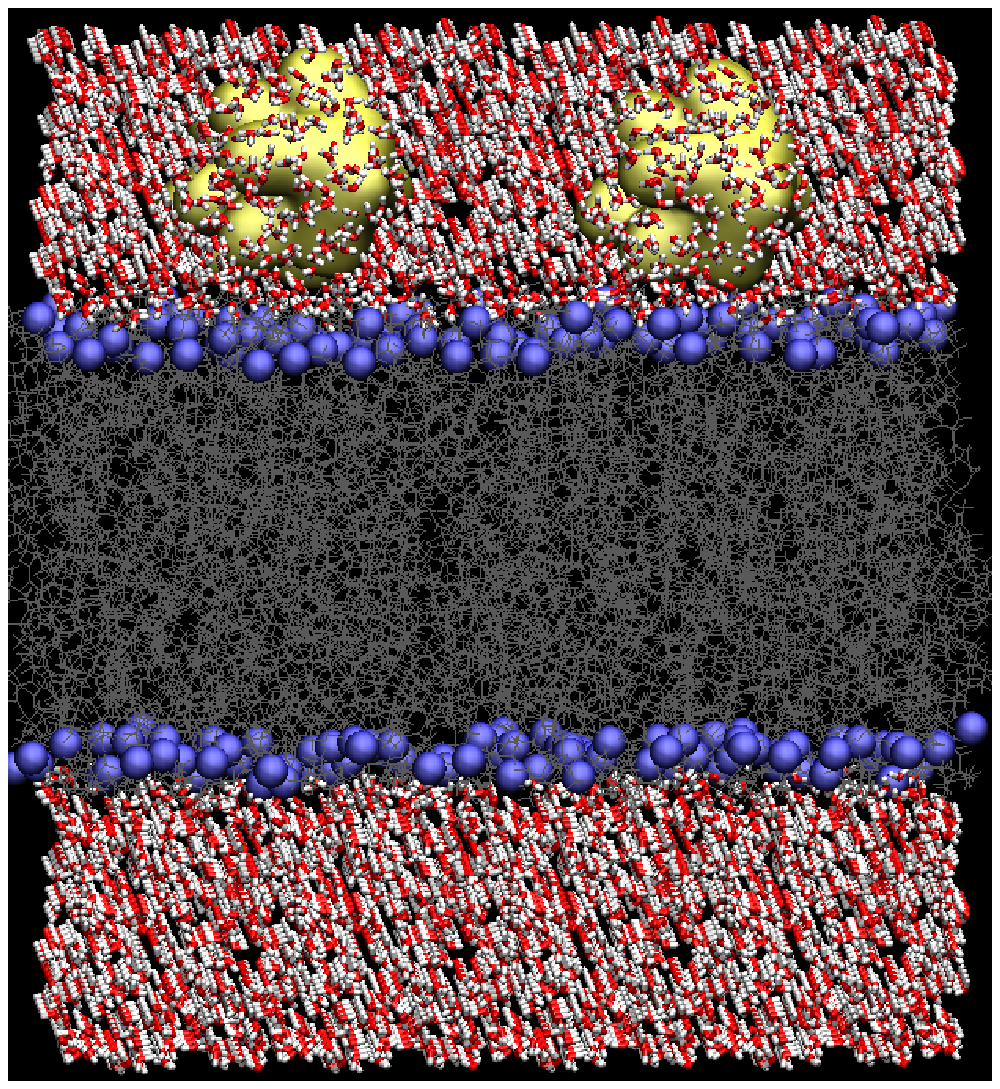}
\includegraphics[width=3cm,height=3.5cm]{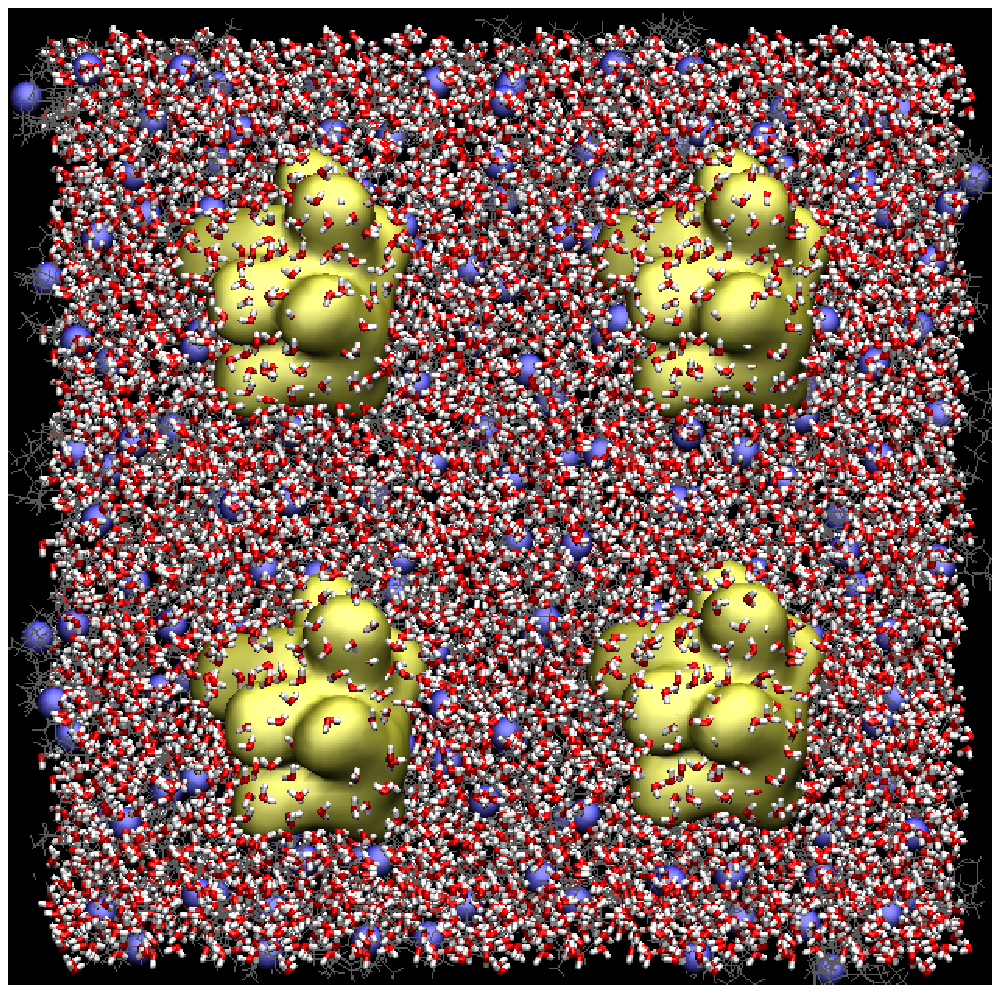}
\includegraphics[width=3cm,height=3.5cm]{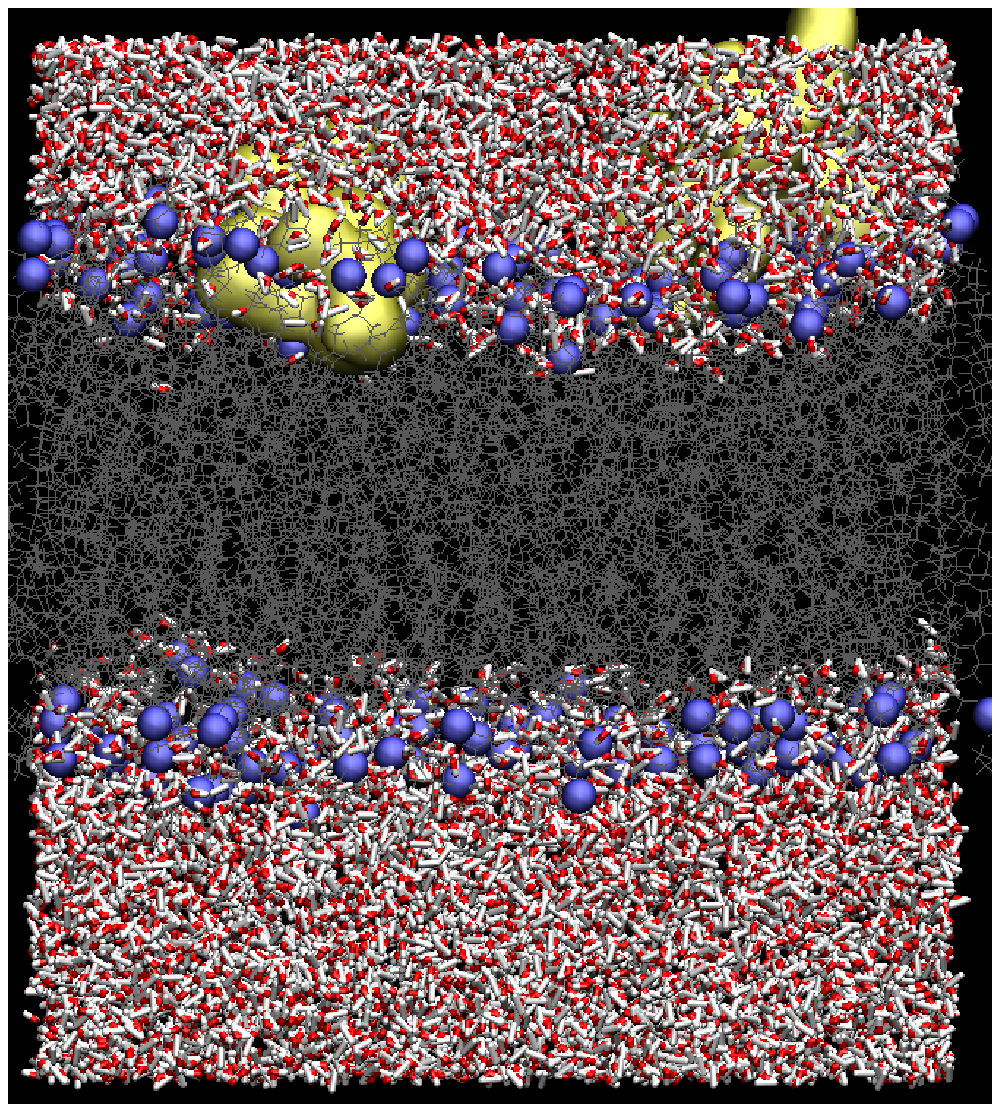}
\includegraphics[width=3cm,height=3.5cm]{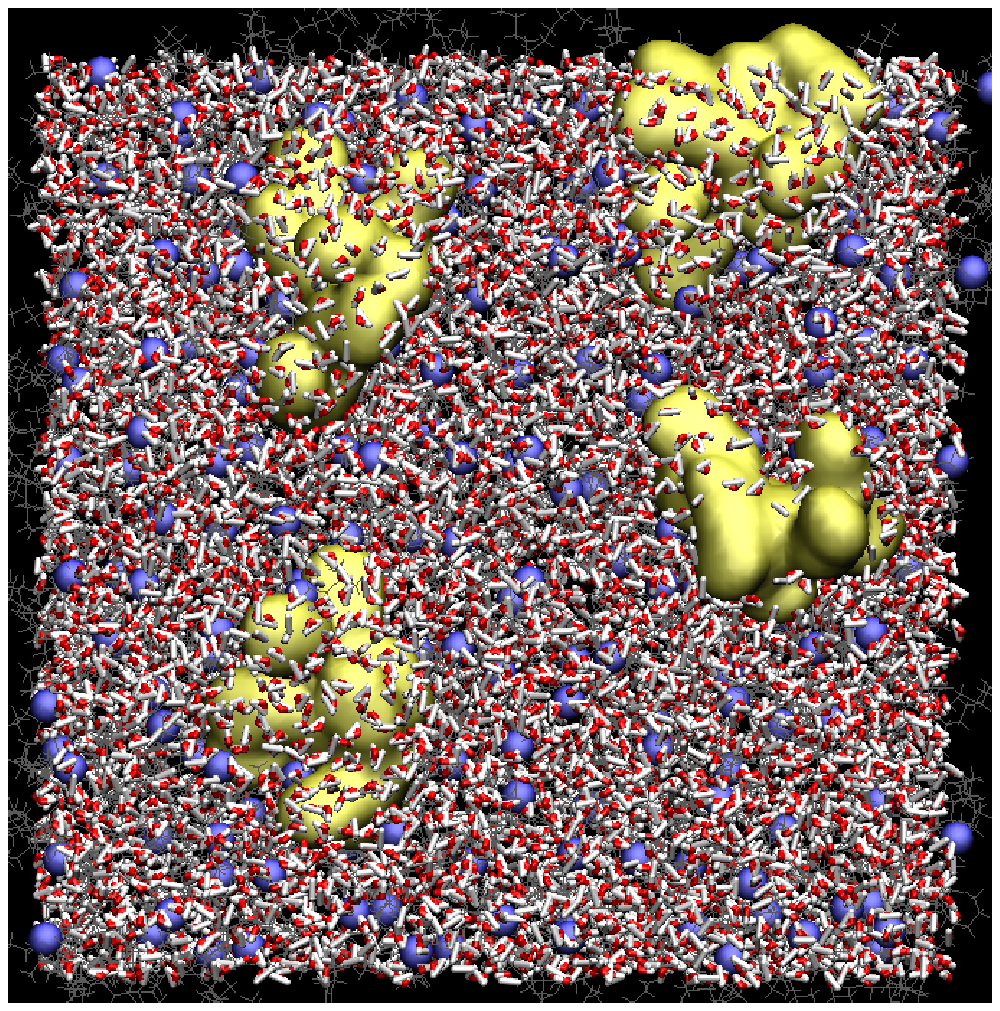}

\caption{The initial setup and a snapshot after equilibration (first low: System I; second low: System II; third low: System III; fourth low: System IV). The yellow is Arg$_9$ (System I, III, IV) or Tat (System II). The blue dots depict phosphorus atoms of DOPC lipids, while the red (orange) dots are those atoms of DOPG(DOPE) lipids. The gray line shows lipid molecules, and water molecules are shown as licorice}
\end{figure*}

\begin{figure*}[ht]
\label{figS2}
\includegraphics[width=8cm,height=8cm]{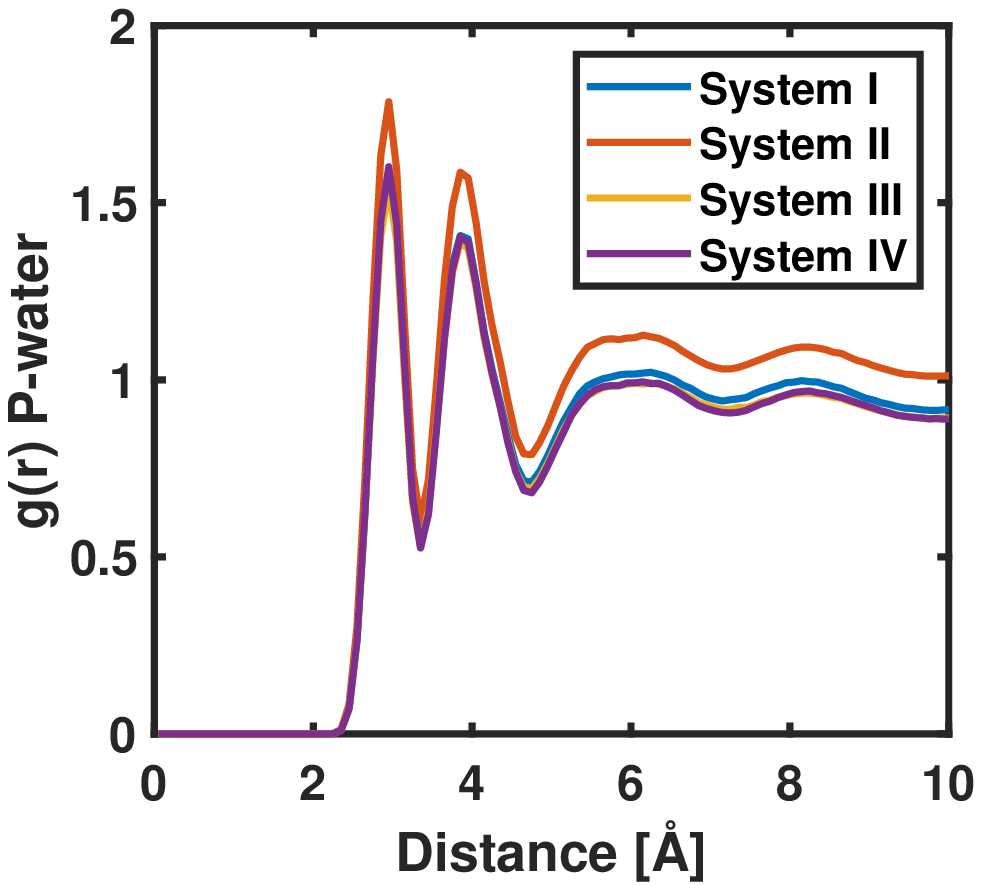}
\includegraphics[width=8cm,height=8cm]{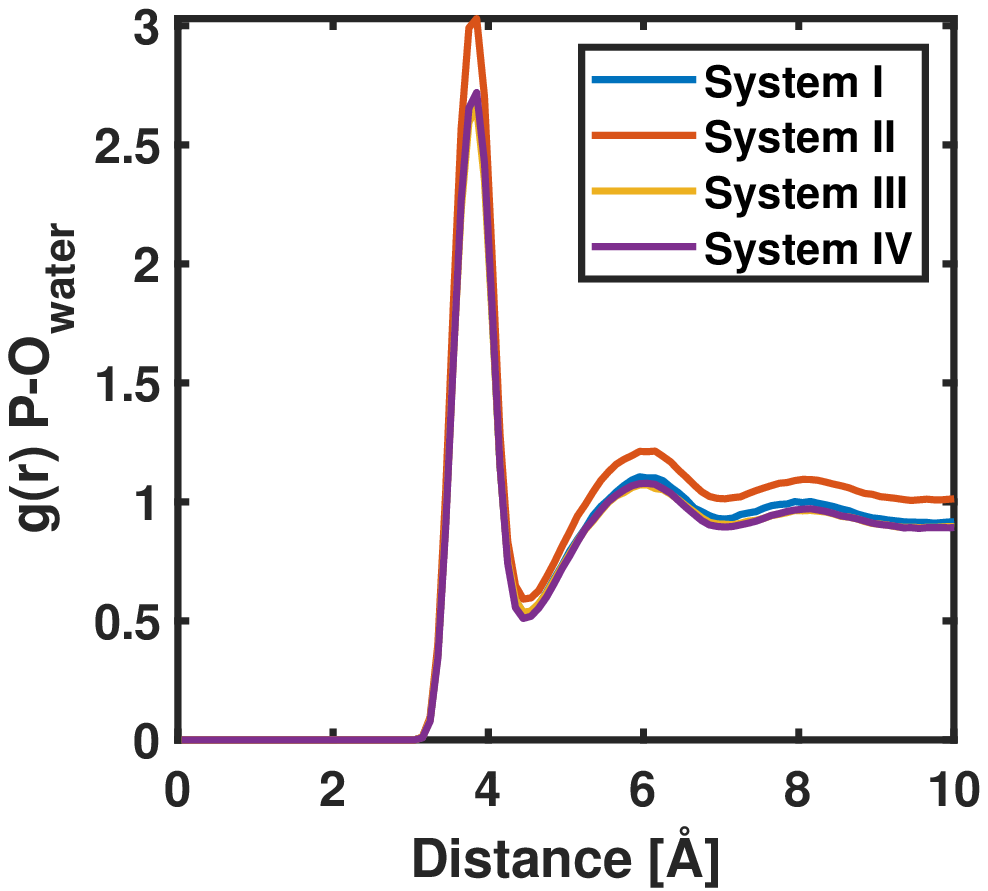}
\includegraphics[width=8cm,height=8cm]{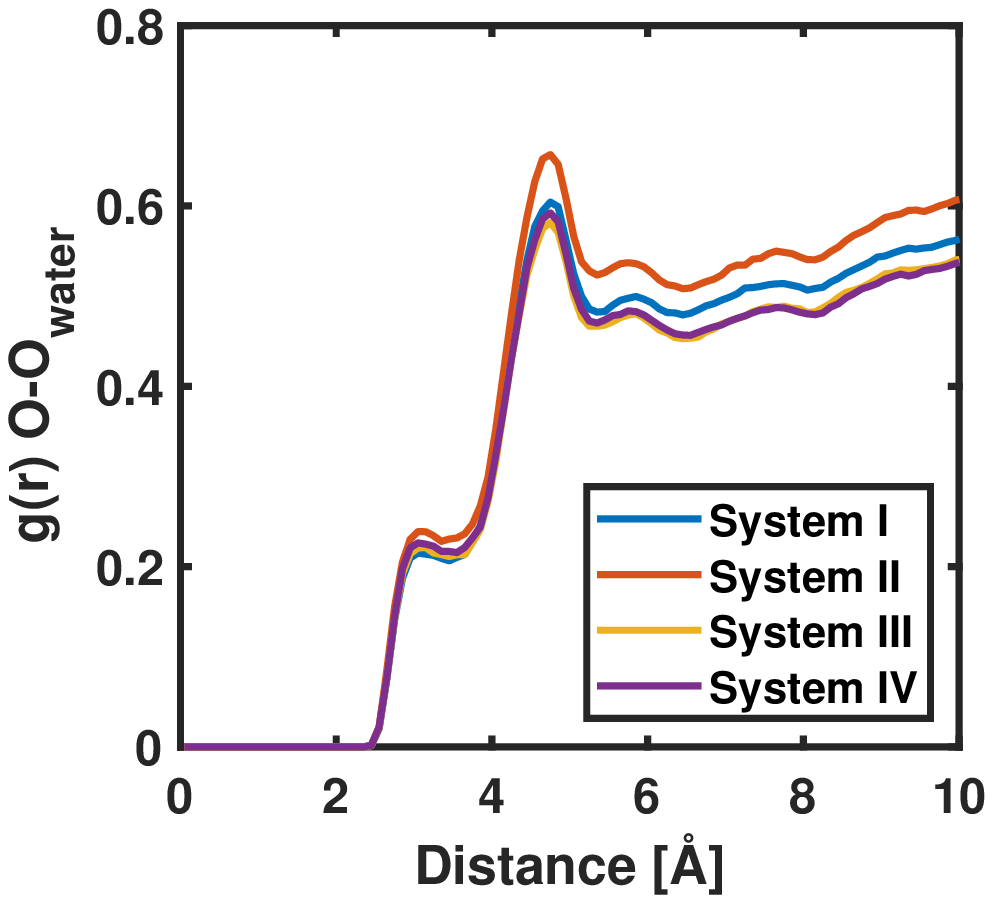}

\caption{Radial distribution function g(r) of (a) phosphorus atoms in lipids vs. water (b) phosphorus atoms vs. oxygen atoms in water molecules (c) ester oxygen in lipids vs. oxygen atoms in water.}
\end{figure*}

\section{Membrane Curvature}

Fig. S3 shows membrane curvature from System I \& II simulations, respectively (left: a side view, right: top view). It indicates significant deformation when Arg$_9$ or Tat penetrates the DOPC/DOPG(4:1) membrane. Each layer (the upper and the lower) consists of only phosphorus atoms in the lipid molecules in this figure. The colored dots on the upper layer depict the $C_\alpha$s of each Arg$_9$ or Tat.

\begin{figure*}[ht]	
\includegraphics[width=16cm,height=7cm]{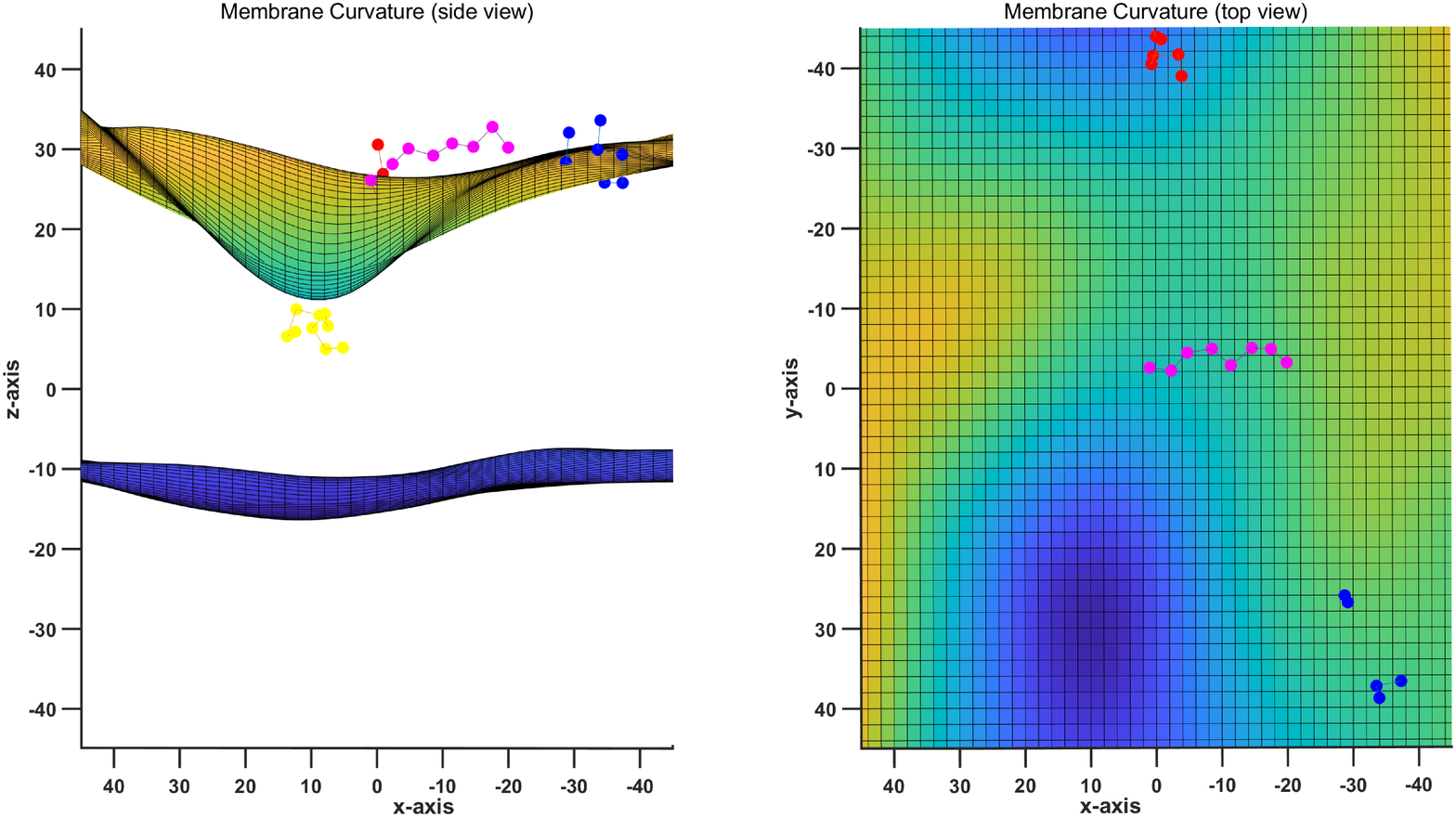}

\includegraphics[width=16cm,height=7cm]{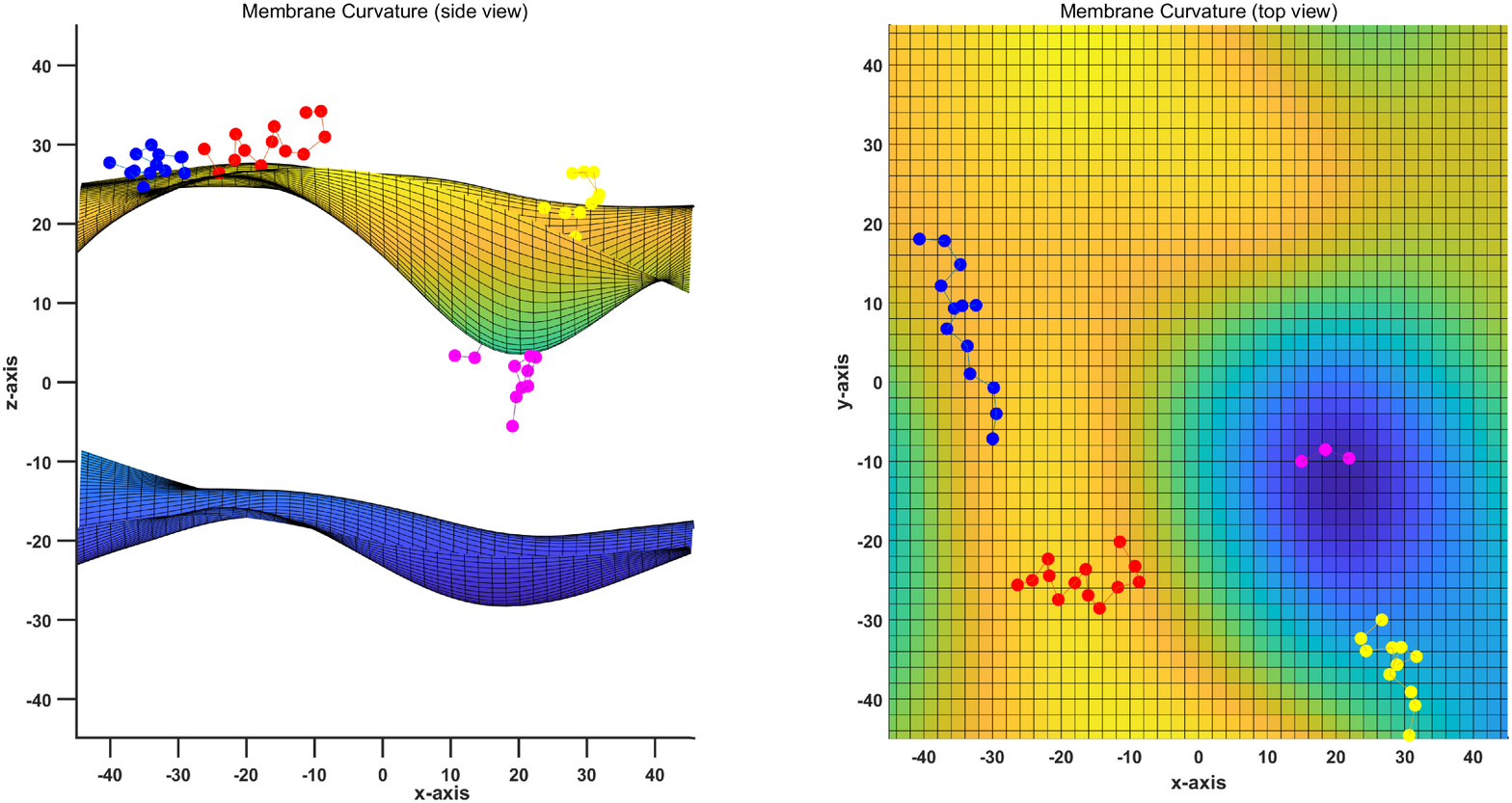}
\caption{Membrane curvature (a) System I (4 Arg$_9$ with DOPC/DOPG(4:1) lipids) (b) System II (4 Tat with DOPC/DOPG(4:1) lipids)}
\end{figure*}

\bibliography{suppl}